\documentclass[aps,prd,reprint,superscriptaddress,floatfix,nofootinbib,preprintnumbers]{revtex4-1}

\usepackage[utf8x]{inputenc}
\usepackage{microtype}

\usepackage{graphicx}
\usepackage[dvipsnames]{xcolor}
\usepackage{rotating}
\usepackage[caption=false,singlelinecheck=false]{subfig}
\usepackage{amsmath}
\usepackage{amssymb}
\usepackage{bbold}
\usepackage{amsfonts,dsfont}
\usepackage{bm}
\usepackage{physics}
\usepackage{mathtools}
\usepackage{slashed}
\usepackage{nccmath}
\usepackage{verbatim}
\usepackage{float}

\usepackage{tabu}
\usepackage{booktabs}
\usepackage{multirow}
\usepackage{array}
\usepackage{enumitem}

\usepackage{url}
\usepackage{xspace}
\usepackage{siunitx}
\usepackage{xfrac}
\usepackage{hyperref}
\usepackage[nameinlink]{cleveref}
\usepackage{appendix}
\usepackage{xifthen}
\usepackage{xcolor}
\hypersetup{
	colorlinks,
	linkcolor={blue!75!black},
	citecolor={blue!75!black},
	urlcolor={blue!75!black}
}

\usepackage[normalem]{ulem}

\crefname{section}{Section\!}{Sections\!}
\crefname{figure}{Figure\!}{Figures\!}
\crefname{equation}{}{}
\crefname{table}{Table\!}{Tables\!}
\crefname{appendix}{Appendix\!}{Appendices\!}

\def\s0#1#2{\mbox{\small{$ \frac{#1}{#2} $}}}
\def\0#1#2{\frac{#1}{#2}}

\newcommand{\orcid}[1]{\href{https://orcid.org/#1}{\includegraphics[height=1.7ex,width=1.7ex]{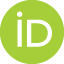}}}

\newcolumntype{x}[1]{>{\centering\arraybackslash\hspace{0pt}}p{#1}}

\DeclareSymbolFont{symbolsC}{U}{pxsyc}{m}{n}

\newcommand{\commentmute}[1]{} 

\graphicspath{{./figures/}}

\begin{document} 
\title{
	$e^+ e^- \to \mu^+ \mu^-$ in the Asymptotically Safe Standard Model}
	
\preprint{DESY-24-208}

	\author{\'Alvaro~Pastor-Guti\'errez~\orcid{0000-0001-5152-3678}}
\affiliation{Max-Planck-Institut f{\"u}r Kernphysik P.O. Box 103980, D 69029, Heidelberg, Germany}
\affiliation{Institut f{\"u}r Theoretische Physik, Universit{\"a}t Heidelberg, Philosophenweg 16, 69120 Heidelberg, Germany}

	\author{Jan M. Pawlowski~\orcid{0000-0003-0003-7180}\,}
\affiliation{Institut f{\"u}r Theoretische Physik, Universit{\"a}t Heidelberg, Philosophenweg 16, 69120 Heidelberg, Germany}
\affiliation{ExtreMe Matter Institute EMMI, GSI Helmholtzzentrum f{\"u}r Schwerionenforschung mbH, Planckstr. 1, 64291 Darmstadt, Germany}

	\author{Manuel~Reichert~\orcid{0000-0003-0736-5726}}
\affiliation{Department of Physics and Astronomy, University of Sussex, Brighton, BN1 9QH, U.K.}

	\author{Giacomo~Ruisi~\orcid{0009-0000-1122-9984}}
\affiliation{Institut f{\"u}r Theoretische Physik, Universit{\"a}t Heidelberg, Philosophenweg 16, 69120 Heidelberg, Germany}
\affiliation{Deutsches Elektronen–Synchrotron DESY, Platanenallee 6, 15738 Zeuthen, Germany}

\begin{abstract}
We study the electron-positron to muon--anti-muon cross-section in the asymptotically safe Standard Model. In particular, we include the graviton contributions to the scattering amplitude, which is computed from momentum-dependent timelike one-particle-irreducible  correlation functions. Specifically, we employ reconstruction techniques for the graviton spectral functions. We find that the full asymptotically safe quantum cross section decreases in the ultraviolet with the centre-of-mass energy, and is compatible with unitarity bounds. Importantly, our findings provide non-trivial evidence for the unitarity of the asymptotically safe Standard Model.
\end{abstract}

\maketitle 

\section{Introduction}
\label{sec:intro}
The unification of general relativity with quantum effects remains one key open question in fundamental high-energy physics. An important contender for such a fundamental theory of quantum gravity is asymptotically safe gravity \cite{Weinberg:1980gg}, where the metric field remains the carrier of the gravitational force. In this purely quantum field theoretical setup, the trans-Planckian ultraviolet (UV) regime of quantum gravity is governed by an interacting fixed point, and gravity is ruled by the same theoretical principles as the Standard Model of particle physics. 

In the past three decades, the field of asymptotically safe gravity has seen substantial progress since the seminal paper \cite{Reuter:1996cp}. By now the existence of asymptotically safe gravity has been put on solid ground; the stability of asymptotic safety has been challenged and tested within a rather extensive class of approximations to the full effective action, ranging from full $f(R)$ approximations~\cite{Kluth:2020bdv, Morris:2022btf, Kluth:2022vnq} over high-order curvature expansion ~\cite{Falls:2020qhj, Knorr:2021slg, Baldazzi:2023pep} to include full momentum dependences of graviton vertices \cite{Christiansen:2015rva, Denz:2016qks, Christiansen:2017bsy, Bonanno:2021squ, Fehre:2021eob}, for recent reviews see \cite{Bonanno:2020bil, Pereira:2019dbn, Reuter:2019byg, Reichert:2020mja, Platania:2020lqb,  Pawlowski:2020qer, Knorr:2022dsx, Platania:2023srt, Pawlowski:2023gym}.

Tremendous progress has been made concerning the interplay between gravity and matter. From studies that investigate the fixed point properties depending on the matter content \cite{Dona:2013qba, Meibohm:2015twa, Eichhorn:2017sok, Eichhorn:2018akn, Eichhorn:2018ydy, Eichhorn:2018nda}, over the impact of graviton fluctuations on the running of matter couplings \cite{Folkerts:2011jz, Eichhorn:2016esv, Eichhorn:2017lry, Christiansen:2017gtg, Eichhorn:2017eht, Pawlowski:2018ixd, Pastor-Gutierrez:2022nki} to the prediction of Standard Model parameters~\cite{Shaposhnikov:2009pv, Eichhorn:2017ylw, Eichhorn:2018whv} and constraining beyond the Standard Model theories~\cite{Reichert:2019car, Eichhorn:2020kca, Eichhorn:2020sbo, Kowalska:2020zve, deBrito:2023ydd}, see \cite{Eichhorn:2022gku} for a review. Recently, the first complete UV-IR trajectories in the full Standard Model including electroweak and quantum gravity threshold effects have been computed \cite{Pastor-Gutierrez:2022nki}.

While the existence of asymptotically safe gravity with and without matter as a stable quantum field theory is well established, the study of observables and its unitarity properties is still in its infancy. It may well be that asymptotically safe theories exist but are not unitary. First studies of unitarity include the computation of spectral functions \cite{Bonanno:2021squ, Fehre:2021eob}, the study of propagator poles \cite{Platania:2020knd, Platania:2022gtt}, first computations of scattering amplitudes \cite{Draper:2020bop, Knorr:2020bjm, Knorr:2022lzn},  and the comparison with positivity bounds \cite{Knorr:2024yiu, Eichhorn:2024wba}.

In this work, we compute the $e^+ e^- \to \mu^+ \mu^-$ scattering cross section as an important probe of observables and of the unitarity of an asymptotically safe theory. It is well known that the leading-order graviton-mediated scattering amplitude grows with the centre-of-mass energy, which violates unitarity in the UV. The full non-perturbative scattering amplitude needs to decrease with the centre-of-mass energy in order to fulfil the Froissart bound \cite{Froissart:1961ux} and to be compatible with unitarity. Scattering in quantum gravity has been computed in effective field theory approaches \cite{Donoghue:1994dn, Donoghue:1993eb, Bjerrum-Bohr:2002gqz} and with eikonal resummations \cite{Giddings:2007qq, Giddings:2007bw, Giddings:2009gj, Giddings:2010pp}. 

The challenge in asymptotically safe quantum gravity has been so far the lack of correlation function at timelike momenta. The functional renormalisation group (fRG), which is the main tool for non-perturbative computation in quantum gravity, is formulated in Euclidean signature and timelike momenta need to be accessed through a Wick rotation \cite{Bonanno:2021squ}. This has recently changed with the development of the spectral renormalisation group \cite{Fehre:2021eob, Braun:2022mgx} that allows for computations directly in backgrounds with Lorentzian signature. Besides the spectral renormalisation group, also other Lorentzian approaches have been developed recently, see \cite{DAngelo:2022vsh, Banerjee:2022xvi, DAngelo:2023tis, DAngelo:2023wje, Banerjee:2024tap, Thiemann:2024vjx, Ferrero:2024rvi, DAngelo:2025yoy}, as well as fRG computations based on an Arnowitt-Deser-Misner (ADM) decomposition, see \cite{Manrique:2011jc, Rechenberger:2012dt, Biemans:2016rvp, Biemans:2017zca, Knorr:2018fdu, Eichhorn:2019ybe, Knorr:2022mvn,  Saueressig:2023tfy, Korver:2024sam, Saueressig:2025ypi}.

Here we exploit two recent developments, the computation of spectral functions \cite{Fehre:2021eob, Bonanno:2021squ} and the computation of UV-IR trajectories in the full Standard Model \cite{Pastor-Gutierrez:2022nki}. This allows us to compute the $e^+ e^- \to \mu^+ \mu^-$ scattering first at leading order and then subsequently improve it to a non-perturbative result by replacing the classical correlation function with the one-particle irreducible (1PI) equivalents from the quantum effective action. 

This work is structured in the following way. In \Cref{sec:LO-scattering}, we derive the leading-order cross-section of $e^+e^- \rightarrow \mu^+\mu^-$ scattering. In \Cref{sec:real-time-quantum-eff-action}, we discuss how we extract 1PI correlation functions from the quantum effective action, and how we access timelike momenta through the graviton spectral function. In \Cref{sec:scattering-ASSM}, we display our result for the cross-section including quantum gravity effects. We also discuss comparisons to other approximation schemes such as renormalisation group (RG) improvements. In \Cref{sec:conclusion}, we summarise our findings.

\begin{figure*}[t]
	\includegraphics[width=1.6 \columnwidth]{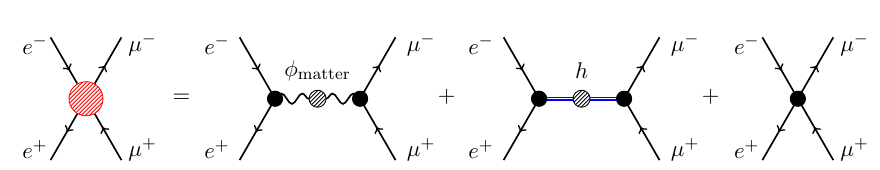} 
	\caption{Electron-positron to muon-antimuon scattering in terms of diagrams with full 1PI vertices. The scattering process is mediated by matter propagators (first diagram) and by a graviton propagator (second diagram) as well as the contact term (third diagram). Full black circles indicate full 1PI vertices and dashed circles indicate full 1PI propagators.}
	\label{fig:scattering-diagrams}
\end{figure*}

\section{$e^+e^- \rightarrow \mu^+\mu^-$ scattering }
\label{sec:LO-scattering}

We begin by computing the leading-order $e^+e^- \rightarrow \mu^+\mu^-$ scattering amplitude from the classical action. We first introduce the classical Einstein-Hilbert and QED action, then derive the leading-order tree-level cross-section from it.

\subsection{Classical action}
\label{sec:classical-action}

The starting point of our computation is the classical Einstein-Hilbert action together with the action of the Standard Model,
\begin{align}
	S = S_\text{EH} + S_\text{SM}\,.
\end{align}
The classical Einstein-Hilbert action reads
\begin{align}\label{eq:EHaction}
	S_\text{EH}= \frac{1}{16 \pi G_\text{N}} \int \! \mathrm d^{4}x \sqrt{g} \left( 2 \Lambda - R \right) + S_\text{gh} + S_\text{gf} \, ,
\end{align}
with the classical Newton constant $G_\text{N}$ and the notation $\sqrt{g} = \sqrt{\vert \det g_{\mu \nu}(x) \vert}$. The classical action is augmented by a standard gauge-fixing and ghost action, see \Cref{app:classical-action} for more details. The latter requires the expansion about a background metric. Here, we work with a flat  Minkowski background $\eta_{\mu \nu}$ and define the fluctuation graviton $h_{\mu\nu}$ with
\begin{align}
	\label{eq:metric-split}
	g_{\mu\nu} = \eta_{\mu\nu} + \sqrt{G_\text{N}} h_{\mu\nu}\,.
\end{align}
For the matter action, we focus on the relevant parts for the computation of the $e^+e^- \rightarrow \mu^+\mu^-$ cross-section. In the SM, this is given by the QED action since the dominant contribution at very high energies is provided by the photon exchange which accounts for the weak and hypercharge sectors, 
\begin{align}
	\label{eq:QED}
	S_\text{SM} = \int \! \mathrm d^4x  \sqrt{g} \left[ -\frac14  F^{\mu\nu} F_{\mu\nu} +  \bar{\psi}_\ell \left( i \slashed{\nabla} - m_\ell   \right) \psi_\ell \right] + \dots \,.
\end{align}
The QED action already includes the interactions between dynamical gravitons and fermions as well as gravitons and photons in a minimal manner. In \cref{eq:QED}, the index $\ell\in(e,\mu)$ labels the flavour of the fermions and $m_\ell$ is their respective mass. In the Standard Model, these masses are generated by the Higgs mechanism. Here, we include this explicit mass term for simplicity and work in the high-energy limit $s\gg m_\ell$.

The $\slashed{\nabla}$ in \cref{eq:QED} indicates the contraction of the spin-covariant derivative $\nabla^{\mu}$ with the Dirac gamma matrices $\gamma_\mu$. The covariant kinetic terms for the fermion fields in~\cref{eq:QED} lead to a minimal coupling between gravity and matter. For the formulation of spinor fields in curved spacetime the spin-base invariance formalism has been introduced~\cite{Weldon:2000fr, Gies:2013noa, Lippoldt:2015cea}. It is based on the space-time dependence of the Dirac matrices required by the general anticommutation relation $\lbrace \gamma_\mu, \gamma_\nu \rbrace = 2 g_{\mu \nu} $. This space-time dependence determines the spin connection. The slashed spin-covariant derivative acting on a spinor field reads
\begin{align}
	\slashed{\nabla} \psi = g_{\mu \nu} \gamma(x)^\mu \nabla^{\nu} \psi = g_{\mu \nu} \gamma(x)^\mu \left( D^\nu + \Gamma(x)^\nu \right) \psi \, ,
\end{align} 
where $\Gamma^\mu$ is the spin connection.

\subsection{Differential cross section}
\label{sec:cross-section}
The differential cross section in the centre-of-mass frame for the process of interest depicted diagrammatically in \Cref{fig:scattering-diagrams} reads 
\begin{align}
	\label{eq:dsCM}
	\left. \frac{\mathrm d\sigma_{\rm tot}}{\mathrm d\Omega} \right|_{\rm CM} = \frac{1}{64 \pi^2 s} \frac{p_\mu}{p_e} \langle \vert \mathcal{M}_{fi} \vert^2 \rangle \,  \theta \left(\sqrt{s} - 2\, m_\mu\right)  ,
\end{align}
where $p_e$ and $p_\mu$ are the momenta of the incoming electron and outgoing muon, $m_\mu$ the mass of the latter and $\sqrt{s}$ the centre-of-mass energy. In~\cref{eq:dsCM}, the averaged matrix element squared $\langle \vert \mathcal{M}_{fi} \vert^2 \rangle$ accounts for all amplitudes depicted on the right-hand side of \Cref{fig:scattering-diagrams}. The abbreviation $\phi_{\rm matter}$ includes all contributing matter fields, namely the photon $\gamma$, the $Z$ boson, and the Higgs $H$, i.e., $\phi_{\rm matter} = (\gamma, Z, H)$. In this work, we are neglecting the $Z$ and $H$ mediated processes. 

At leading order, the contact term in \Cref{fig:scattering-diagrams} does not contribute and we are left with the photon (${\cal M}_\gamma$) and graviton (${\cal M}_h$) mediated scattering processes. The total averaged matrix element squared is given by
\begin{align}
	\label{eq:matrix-element}
	\langle \vert \mathcal{M}_{fi} \vert^2 \rangle = \langle \vert \mathcal{M}_\gamma + \mathcal{M}_h \vert^2 \rangle\,.
\end{align}
Beyond leading order, quantum gravity effects modify both matrix elements in \cref{eq:matrix-element} and additionally the third diagram in \Cref{fig:scattering-diagrams} becomes relevant. Indeed all vertices and propagators in \Cref{fig:scattering-diagrams} are modified by graviton loops, which are negligible below the Planck scale because of the smallness of Newton's coupling in this regime. These low energy quantum corrections have been extensively studied by means of effective field theory techniques~\cite{Donoghue:1994dn, Donoghue:1993eb, Bjerrum-Bohr:2002gqz}. At leading order, the dominant contribution above the Planck scale stems from the matrix element of the graviton-mediated scattering,
\begin{align}\label{eq:MEgraviton}
	\mathrm{i} \mathcal{M}_{h} = \text{J}^{(e^{-} e^{ +})} \,  (S^{(hh)})^{-1} \, \text{J}^{(\mu^- \mu^+)}.
\end{align}
Here we are suppressing the space-time indices of the graviton two-point function and the currents. They are explicitly given in \Cref{app:matrix-element}. The fermion currents in \cref{eq:MEgraviton} read
\begin{align}\label{eq:currents}
	\text{J}^{(e^- e^+)}   & =  \bar{v}(p_{\text{e}^+})\, S^{(h\, e^- e^+)}\, u(p_{\text{e}^-})\,, \notag  \\ 
	\text{J}^{(\mu^- \mu^+)} & = \bar{u}(p_{\mu^-})\, S^{(h \, \mu^- \mu^+)}\, v(p_{\mu^+}) \,,
\end{align}
and contain the graviton-electron-positron $ S^{(h\, e^- e^+)}$ and graviton-muon-antimuon $S^{(h \, \mu^- \mu^+)}$ vertices. Note that the graviton propagator depends on the gauge-fixing parameters but only the physical on-shell degrees of freedom contribute to the matrix element, which is independent of the gauge-fixing parameters.

In \cref{sec:scattering-ASSM}, we will go beyond the leading-order result by upgrading the classical vertices by fully momentum-dependent 1PI vertices from the quantum effective action. This will be achieved by employing a vertex expansion for the effective average action, see \cite{Pawlowski:2020qer, Pawlowski:2023gym}, and by utilising recent results that allow the computation of non-perturbative propagators and vertices on backgrounds with Lorentzian signature \cite{Bonanno:2021squ, Fehre:2021eob}.

\subsection{Matrix elements at leading order}
\label{sec:matrix-element}
There are two contributions to the differential cross section \cref{eq:dsCM} at leading order: the graviton and the photon mediated diagrams, both depicted in \Cref{fig:scattering-diagrams}. Here, we compute the tree-level diagrams, which will be the basis for the full scattering amplitude with dressed quantum vertices later on.

Let us first consider the graviton-mediated part, where detailed computation is given in \cref{app:matrix-element}. We take into account the dependence of the matrix element on the masses of the external on-shell states and confirmed the gauge invariance of the resulting matrix element. In the relativistic limit, $\langle \vert \mathcal{M}_h \vert^2 \rangle$ is highly simplified and can be expressed in the centre of mass frame solely as a function of $\sqrt{s}$ and the scattering angle $\theta$,
\begin{align}
	\label{eq:Mh-leading}
	\langle \vert \mathcal{M}_h \vert^2 \rangle =  \pi^2 s^2 G_\text{N}^2 ( 1 - 3 \cos^2 \theta + 4 \cos^4 \theta )\,.
\end{align}
This matrix element is dominated by the transverse-traceless (TT) mode of the graviton propagator. The scalar mode of the graviton gives a subleading contribution that vanishes identically in the high-energy limit since the scalar mode only mixes with the fermion mass terms when taking the trace. This feature allows us to focus on the transverse-traceless mode when computing the full quantum propagator.

The next contribution is the well-known photon mediated scattering with the averaged matrix element squared 
\begin{align}
	\label{eq:Mgamma-leading}
	\langle \vert \mathcal{M}_\gamma \vert^2 \rangle =  16 \pi^2 \alpha_e^2 (1 + \cos^2 \theta)\,,
\end{align}
where $\alpha_e = e^2/4\pi$ and 
\begin{align}\label{eq:ecoupling}
	e = g \, \sin \theta_w = \frac{g \, g^\prime}{\sqrt{g^2 + g^{\prime 2}}}\,,
\end{align}
with $g \equiv g_2$ and $g^{\prime} \equiv g_Y \equiv \sqrt{3/5} \, g_1$ being the weak isospin and the weak hypercharge couplings, respectively. 

With  both matrix elements at hand, we can derive the interference term 
\begin{align}
	\langle \vert \mathcal{M}_{\gamma}^\ast \mathcal{M}_h + \mathcal{M}_{h}^\ast \mathcal{M}_{\gamma} \vert \rangle = \alpha_e G_{\text{N}} s \cos^3 \theta\,,
\end{align}
which renders the following total differential cross-section at leading order
\begin{align}\label{eq:dsigmaTOT}
	\frac{\text{d} \sigma_{\rm tot } }{\text{d} \Omega}  = & \frac{\alpha_e^2}{4  s} (1 + \cos^2 \theta)  + \frac{G_\text{N}\, \alpha_e}{4} \cos^3 \theta  \notag \\
	& + \frac{G_\text{N}^2  }{64}s  (1 - 3 \cos^2 \theta + 4 \cos^4 \theta).
\end{align}
The interference terms show a $\cos^3 \theta$ dependence, with $\theta$ being the scattering angle in the centre-of-mass frame. Since odd powers vanish when integrating over the total solid angle $\Omega$, the interference terms to do contribute to the total leading-order cross-section. In summary, the total leading-order cross-section is given by,
\begin{align}\label{eq:sigmaTOT}
	\sigma_{\rm tot }(s) = \sigma_{\gamma}(s)+ \sigma_{h}(s)= \frac{4 \pi \alpha_e^2}{3 s} +  \frac{\pi G_{\text{N}}^2 \, s}{20}\,,
\end{align}
where the photon-mediated contribution scales with $1/s$, while the graviton-mediated contribution increases with $s$. The latter is rooted in the negative mass dimension of the Newton coupling in four dimensions and reflects that the classical cross-section violates unitarity bounds at leading order.

\section{Real-time correlation functions from the quantum effective action}
\label{sec:real-time-quantum-eff-action}

In this section, we access the real-time domain of correlation functions via the spectral representation of the propagator.  The determination of timelike correlation functions from their Euclidean counterparts is a tangled and challenging task common to non-perturbative methods. In quantum gravity, this task is even more difficult, as we lack a proper definition of the Wick rotation even on conceptual grounds, see also \cite{Baldazzi:2018mtl, Baldazzi:2019kim}. These issues have been addressed successfully for the first time in \cite{Bonanno:2021squ} by assuming the existence of a Wick rotation and using reconstruction techniques. In \cite{Fehre:2021eob}, a Lorentzian spectral functional RG has been set up. This approach has been used for the first direct Lorentzian computation of the graviton propagator, drawing from the spectral fRG and Dyson-Schwinger approach put forward in  \cite{Horak:2020eng, Horak:2021pfr, Horak:2022myj, Braun:2022mgx} for quantum field theories. Here we review and build upon these results and utilize them for the computation of the scattering amplitude.

In recent years, further Lorentzian approaches have been developed, see \cite{DAngelo:2022vsh, Banerjee:2022xvi, DAngelo:2023tis, DAngelo:2023wje, Banerjee:2024tap, Thiemann:2024vjx, Ferrero:2024rvi, DAngelo:2025yoy}, with some of them closely related to the present one. Furthermore, fRG approaches based on ADM-type decompositions have been considered in e.g.~\cite{Manrique:2011jc, Rechenberger:2012dt, Biemans:2016rvp, Biemans:2017zca, Knorr:2018fdu, Eichhorn:2019ybe, Knorr:2022mvn,  Saueressig:2023tfy, Korver:2024sam, Saueressig:2025ypi}. Given the subtle nature of the background independence of Lorentzian approaches, explicit results for Lorentzian propagators and vertices from other Lorentzian approaches than the spectral one used here are much wanted for and are to be expected in the near future.

\begin{figure*}[htbp]
	\includegraphics[scale=.9]{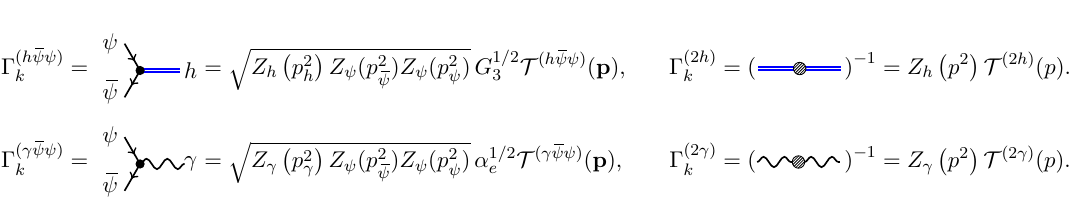}
	\caption{ Vertex dressing of the \textit{n}-point functions used in this work for the computation of the cross section. The vertex dressing consists of the respective wave function renormalisations, couplings, and tensor structures. The first line displays the graviton-fermion interaction vertex and the graviton two-point function (inverse of the graviton propagator). The second line shows the photon-fermion interaction vertex and the photon two-point function.}
	\label{fig:dressed}
\end{figure*}

\subsection{Quantum effective action}
\label{sec:eff-act}

The quantum effective action is the generating functional of 1PI correlation functions including all quantum effects. These are the correlation functions that we want to access and employ in the cross-section computation displayed in \Cref{fig:scattering-diagrams}.

The functional renormalisation group (fRG) is a convenient tool for the computation of the quantum effective action. In this approach, an infrared (IR) regulator is introduced in the path integral, which suppresses quantum fluctuations below a given IR cutoff scale $k$, leading to the respective scale-dependent effective action $\Gamma_k[\phi]$. Then, the full effective action $\Gamma[\phi]=\Gamma_{k=0}[\phi]$ is obtained by successively integrating out momentum fluctuations at the scale $k$. The respective flow equation for $\Gamma_k$, the Wetterich equation \cite{Wetterich:1992yh, Morris:1993qb, Ellwanger:1993mw}, reads
\begin{align}\label{eq:FlowEq}
	\partial_t \Gamma_k \left[\phi \right] &= \frac{1}{2} \text{Tr}\!\left[\frac{1}{\Gamma_k^{(2)} +R_k } \partial_t R_k \right] ,
\end{align}
where 
\begin{align}\label{eq:DefGn}
	\Gamma^{(n)}_{\phi_{i_1}\cdots \phi_{i_n}}[\phi](p_1,...,p_n) =\frac{\delta\,\Gamma[\phi]}{\delta \phi_{i_1}(p_1)\cdots \delta\phi_{i_n}(p_n)}\,. 
\end{align}
For a recent review on the fRG we refer to \cite{Dupuis:2020fhh}. In the context of scattering amplitudes we are interested in the computation of the graviton and photon propagator, the fermion-graviton and fermion-photon vertices since these directly contribute to the scattering amplitude, see \cref{fig:scattering-diagrams}. In this work, we neglect the contribution of the four-fermion interaction.

In summary, we are using the $n$-point functions
\begin{align}
	\label{eq:computed-n-point-fcts}
	\Gamma^{(hh)} ,\, \Gamma^{(\gamma\gamma)} ,\, \Gamma^{(h\bar\psi\psi)} ,\, \Gamma^{(\gamma\bar\psi\psi)}\,.
\end{align}
The flow equations for these $n$-point functions depend on correlation functions of the order $n+2$, which creates an infinite tower of coupled flow equations. The system is then solved by truncating the expansion at a given order, for a review of the approach see \cite{Pawlowski:2020qer, Pawlowski:2023gym}. The momentum-dependent $n$-point functions given in \cref{eq:computed-n-point-fcts} have been previously computed in: $\Gamma^{(hh)}$ \cite{Christiansen:2012rx, Christiansen:2014raa, Knorr:2021niv, Bonanno:2021squ, Fehre:2021eob}, $\Gamma^{(\gamma\gamma)}$ \cite{Folkerts:2011jz, Christiansen:2017cxa}, $\Gamma^{(h\bar\psi\psi)}$ \cite{Meibohm:2015twa, Eichhorn:2018nda},  $\Gamma^{(\gamma\bar\psi\psi)}$ \cite{Pastor-Gutierrez:2022nki}. Additionally, we are using the graviton three-point function $\Gamma^{(3h)}$ computed in \cite{Christiansen:2015rva, Denz:2016qks}. Here we build upon these results and utilise them for the computation of the scattering amplitude.

We define the RG-invariant $n$-point correlation functions $\bar \Gamma^{(n)}$ with
\begin{align}
	\label{eq:def-gamma-bar}
	\Gamma^{(\phi_1 ... \phi_n)}_k (\mathbf p ) = \left( \prod_i \sqrt{Z_{\phi_i}(p_i^2)} \right) \bar \Gamma^{(\phi_1 ... \phi_n)}_k (\mathbf p ) 
\end{align}
where $\mathbf p = (p_1,\dots,p_n)$ contains the four-momenta of all $n$ fields of which one can be eliminated using momentum conservation. The $Z_{\phi_i}$ are the momentum-dependent wave function renormalisations of the field $\phi_i$. By construction, the wave function renormalisations cancel out in the computation of the scattering amplitude given in \cref{fig:scattering-diagrams}. Note that they still contribute to the running of the $n$-point functions through the momentum-dependent anomalous dimension
\begin{align}
	\label{eq:anom-dim}
	\eta_{\phi_i} (p^2) = - \partial_t \ln Z_{\phi_i} (p^2)\,.
\end{align}
The  RG-invariant $n$-point correlation functions are parametrised with 
\begin{align}
	\label{eq:dressing-gamma-bar}
	\bar \Gamma^{(\phi_1 ... \phi_n)}_k (\mathbf p )  = \sum_j \mathcal{A}^{(n)}_j (\mathbf p)\, \mathcal{T}_j^{(\phi_1 ... \phi_n)}(\mathbf p)\,.
\end{align}
The index $j$ labels the different tensor structures $\mathcal T_j$ of a given $n$-point function and the $\mathcal A_j$ contain the couplings that describe the running of the correlation function.

The graviton propagator is parameterised by the wave function renormalisation and the graviton mass parameter $\mu$, which is related to the cosmological constant. After rescaling with the wave function renormalisation, the RG-invariant propagator is given by (without regulator)
\begin{align}
	\label{eq:graviton-prop}
	\mathcal G_{hh,\mu\nu\rho\sigma} (p) = \frac{32\pi}{p^2+\mu k^2} \mathcal T_{tt,\mu\nu\rho\sigma}(p) -\frac{16\pi}{p^2+\mu k^2} \mathcal T_{0,\mu\nu\rho\sigma}(p)\,,
\end{align}
where we have used $\beta =1$ and the Landau limit of the gauge fixing parameter, $\alpha \to 0$. The latter ensures that the graviton propagator is fully described by the transverse-traceless and a scalar mode. For other gauge fixing parameters, see \Cref{app:propagator}. In our work, we are identifying $ Z_{h_{0}} = Z_{h_{tt}}$ since the scalar mode is not contributing to the scattering amplitude in the high-energy limit at leading order.

For the graviton-fermion vertex, a convenient choice for the parameterisation of the coupling is 
\begin{align}
	\label{eq:fermion-graviton-newton}
	\mathcal{A}^{(h_{tt}\bar\psi\psi)} (\mathbf p) = \sqrt{G_{\text{N},h\bar\psi\psi} (\mathbf p)} \,,
\end{align}
where for the graviton leg we take the transverse-traceless tensor structure. In the IR, the coupling is identical to the classical Newton coupling $G_{\text{N},h_{tt}\bar\psi\psi} (\mathbf p \to 0) = G_\text{N}$. For finite momenta, the coupling encompasses contributions from higher-order operators. Due to these properties, this coupling is called an avatar of the Newton coupling \cite{Eichhorn:2018akn, Eichhorn:2018ydy}. Note that there is only one avatar for both, the electron-graviton vertex and the muon-graviton vertex. Due to the universality of gravity, the couplings are identical for the relevant scales considered here.

Another important avatar of the Newton coupling is that of the transverse-traceless three-graviton vertex
\begin{align}
	\label{eq:three-graviton-newton}
	\mathcal{A}^{(3h_{tt})} (\mathbf p) = \sqrt{G_{\text{N},3h} (\mathbf p)} \,,
\end{align}
which also has the property to match the classical Newton coupling in the IR, $G_{\text{N},h\bar\psi\psi} (\mathbf p \to 0) = G_\text{N}$. Note that in \cref{eq:three-graviton-newton} we have suppressed that there are multiple tensor structures in the transverse-traceless three-graviton vertex, and we have picked out one of them.

The two avatars of the Newton coupling match at small momentum, but it was also found that their behaviour at the momentum symmetric point at the UV fixed point is approximately identical. In this regime, the couplings are related by a Ward identity and the non-trivial approximate equality was named effective universality \cite{Eichhorn:2018akn, Eichhorn:2018ydy}.

In straight analogy to the graviton vertices, the photon-fermion vertex is parameterised with
\begin{align}
	\label{eq:fermion-photon}
	\mathcal{A}^{(\gamma \bar\psi\psi)} (\mathbf p) = \sqrt{\alpha_e (\mathbf p)} \,,
\end{align}
Similarly to the avatars of the Newton coupling, this coupling matches its classical version for small momenta, $\alpha_e (\mathbf p \to 0) = \alpha_e$. In the UV, this coupling includes quantum gravity effects, which correspond to graviton loops contributing to the photon propagator and to the photon-fermion vertex. All vertex dressings are summarised in \Cref{fig:dressed}.

Each of the computed correlation functions in \cref{eq:computed-n-point-fcts} can be expressed in terms of form factors of the quantum effective action, see \cite{Bosma:2019aiu, Knorr:2019atm, Draper:2020bop, Draper:2020knh, Knorr:2022lzn, Knorr:2022dsx} for recent works on the form factor approach in quantum gravity. To illustrate the correspondence between the approaches, we provide shortly the translation from the correlation functions to the form factors, see also \cite{Knorr:2021niv, Pawlowski:2023gym, Pawlowski:2023dda}. The graviton propagator is described by the effective action
\begin{align}
	\Gamma_{hh} &=  \int \mathrm d^4x \sqrt{g}\left( \frac{R}{16 \pi G_\text{N}} + C_{\mu\nu\rho\sigma} f_C(\Box) C^{\mu\nu\rho\sigma} \right) .
\end{align}
Here we have suppressed the second form factor $R f_R(\Box)R$ since it does not contribute to the transverse-traceless graviton propagator. This is the full set of form factors contributing to the graviton propagator around flat Minkowski space.

The photon propagator is described by the effective action
\begin{align}
	\Gamma_{\gamma\gamma} = \int \! \mathrm d^4x  \sqrt{g} \left[ -\frac14  F^{\mu\nu} f_F(\Box) F_{\mu\nu}  \right] .
\end{align}
The form factors $f_F(\Box)$ and $f_C(\Box)$ directly relate to the wave function renormalisations $Z_\gamma(p^2)$ and $Z_{h_{tt}}(p^2)$, respectively.

The fermion-graviton vertex is described by the effective action
\begin{align}
	\Gamma_{h\bar\psi\psi}  = \int \! \mathrm d^4x  \sqrt{g} \left[ f_{R\bar\psi\psi}(\nabla_1,\nabla_2,\nabla_3) R^{\mu\nu} \bar\psi \gamma_\mu \nabla_\nu \psi \right],
\end{align}
where $\nabla_1$ only acts upon $R_{\mu\nu}$, $\nabla_2$ only on $\bar\psi $, and $\nabla_3$ only on $\psi$. The form factor $f_{R\bar\psi\psi}$ is directly related to 	$\mathcal{A}^{(h_{tt}\bar\psi\psi)} (\mathbf p)$. We again suppressed other graviton-fermion operators such as $R \bar\psi \slashed{\nabla} \psi$ since they do not contribute to the vertex with a transverse-traceless graviton.

Last, the fermion-photon vertex is described by
\begin{align}
	\Gamma_{\gamma\bar\psi\psi} = \int \! \mathrm d^4x  \sqrt{g} \left[  f_{\bar\psi\psi \gamma}(\nabla_1,\nabla_2,\nabla_3) \bar{\psi} i \slashed{\nabla}\psi \right],
\end{align}
where the derivatives of the form factor $f_{\bar\psi\psi \gamma}$ act upon $\bar\psi$, $A_\mu$, and $\psi$, respectively, and the form factor is directly related to $\mathcal{A}^{(\gamma\bar\psi\psi)} (\mathbf p)$.

\subsection{Graviton spectral function}
\label{sec:spectral-functions}

\begin{figure*}[tbp]
	\includegraphics[width=.48\linewidth]{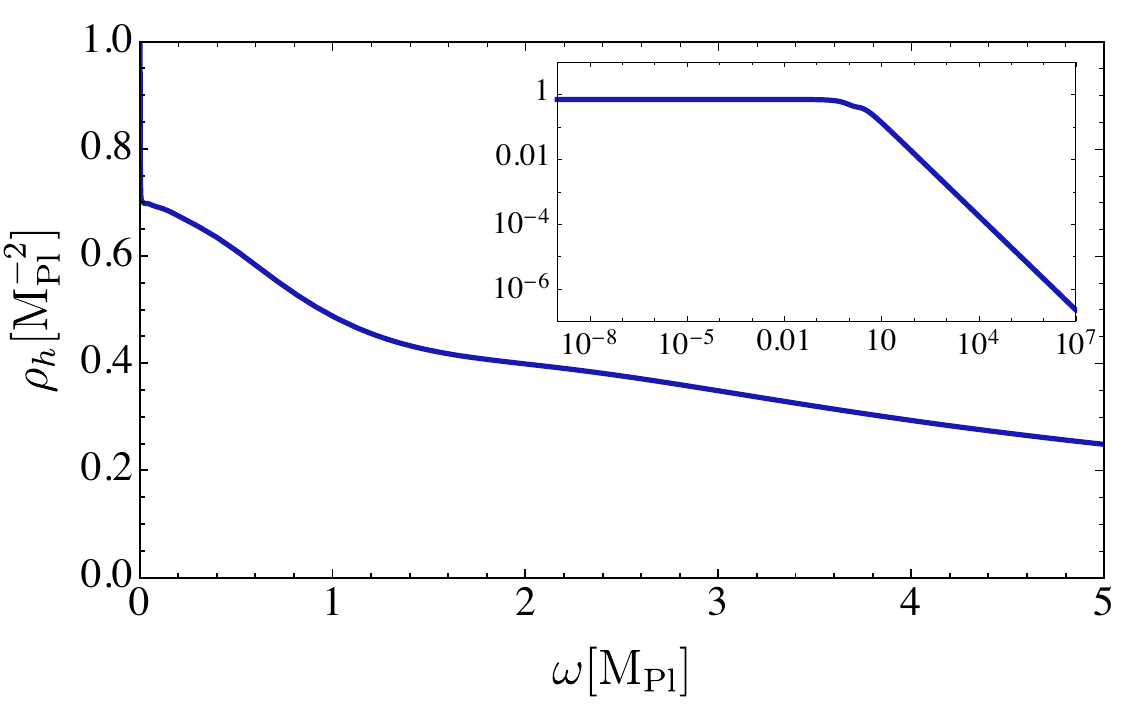}
	\hfill
	\includegraphics[width=.48\linewidth]{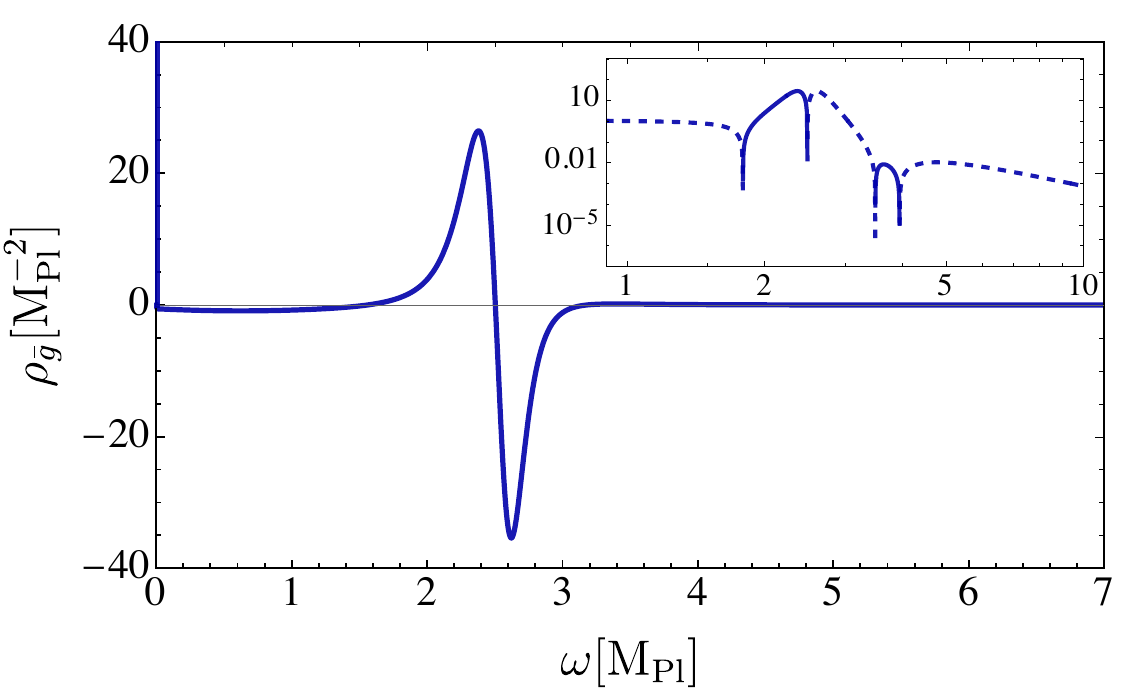} 
	\caption{Spectral function of the fluctuation (left panel) and background (right panel) graviton fields. Both feature a Dirac delta peak at $\omega = 0$ (massless graviton) and a smooth multi-particle continuum for positive frequencies. The background spectral function contains positive and negative parts and has a vanishing spectral weight. The latter feature is better displayed in the inset plot in the top-right corner, in which we show the absolute value. The solid and dashed lines in the inlay show the positive and negative parts of the spectral function, respectively.}
	\label{fig:spectral-function}
\end{figure*}

We use the spectral representation of the graviton propagator to access the timelike momenta that enter the scattering amplitude. We build upon the techniques introduced in \cite{Bonanno:2021squ, Fehre:2021eob}, which we review here.

We assume that propagators have a K{\"a}llen-Lehmann (KL) spectral representation~\cite{Kallen:1952zz, Lehmann1954}. This spectral representation serves as a bridge connecting spacelike (Euclidean) and timelike physics. The existence of the spectral function offers access to the propagator for general complex momenta, in particular for timelike momenta as relevant for graviton-mediated scattering processes.

In momentum space, the time-ordered propagators of physical fields (asymptotic states) are related to their spectral representation $\rho$ by
\begin{align}
	\label{eq:MinkSpecInt}
	\mathcal{G} (p_0, \vert \vec{p} \vert) = \int^{\infty}_{0} \frac{\mathrm{d} \omega}{\pi} \frac{\omega \, \rho(\omega, \vert \vec{p} \vert)}{p_0^{2} - \omega^2 + i \epsilon} ,
\end{align}
with the temporal and spatial momentum $p_0$ and $\vec{p}$, respectively, and the spectral values $\omega$.

The KL representation in~\cref{eq:MinkSpecInt} provides the full propagator in terms of a spectral integral over on-shell propagators $1/(p_0^2 - \omega^2 + i \epsilon)$ of states with pole masses $\omega$. In this work, the reconstruction of the spectral function is done with the Euclidean propagator $\mathcal{G}_{\text{E}}(p_0)$ for Euclidean momenta $p_0$, which is related to the spectral function $\rho$ according to
\begin{align}
	\label{eq:SpecIntEucl}
	\mathcal{G}_{\text{E}}(p_0) = \int_{0}^{\infty} \frac{\mathrm{d} \omega}{\pi} \frac{\omega \, \rho(\omega, \vert \vec{p} \vert)}{\omega^2 + p_0^2}.
\end{align}
Equivalently, the spectral function can be obtained from the Euclidean propagator using an analytic continuation into the complex plane,
\begin{align}
	\label{eq:ImGE}
	\rho(\omega, \vert \vec{p} \vert) = \lim_{\epsilon \rightarrow 0} 2 \, \text{Im} \, \mathcal{G}_{\text{E}}\, (\, p_0 = -i(\omega + i \epsilon)\, , \, \vert \vec{p} \vert  \, ) .
\end{align}
The last equation tells us that $\rho$ acts as a linear response function of the two-point correlator, encoding the energy spectrum of the theory. For asymptotic states, it can be seen as a probability density for the transition to an excited state with energy $\omega$.

The spectral reconstruction procedure used to obtain the real-time propagator for the fluctuation graviton from Euclidean data is described in~\cref{app:reconstruction}. In the present work, we implemented a $1/p^2$ and the hypergeometric function $\mathcal{U}_{1,1}(p^2)$ to reproduce the IR asymptotics featured by the Euclidean fluctuation graviton propagator. In contrast with the previous work~\cite{Bonanno:2021squ}, we made use of the Schlessinger point method (SPM), described in~\cref{app:SPM}, to fit both the UV asymptotics and the transition behaviour from the infrared to the ultraviolet regime.

We assume a uniform shape of the graviton spectral function for all modes and extract it from the TT part. This approximation is justified since this mode is the most dominant contribution in the propagator and, in fact, only the TT mode contributes to the scattering amplitude at leading order in the high-energy limit, see \Cref{sec:matrix-element}.

In \Cref{fig:spectral-function}, we show the spectral functions for the TT-mode of the fluctuation and background graviton fields, which have been respectively computed in this work and \cite{Bonanno:2021squ}. 

Both spectral functions are characterised by a Dirac delta for vanishing frequencies $\omega=0$. This delta stems from the classical part $1/p^2$ of the propagators corresponding to a massless on-shell graviton. Other than that, they have significantly different properties for positive frequencies. The fluctuation graviton spectral function $\rho_h(\omega)$ is always a positive valued function, as it obeys the following spectral sum rule
\begin{align}\label{eq:SpectralSumRule}
	\int_{0}^{\infty} \frac{\text{d}\omega}{\pi} \omega \rho_h(\omega) = \infty,
\end{align}
and this indicates the fluctuation graviton does propagate and mediate the scattering.

In contrast, the background spectral function has a vanishing spectral weight
\begin{align}\label{eq:SpectralSumRule2}
	\int_{0}^{\infty} \frac{\text{d}\omega}{\pi} \omega \rho_{\bar{g}}(\omega) = 0.
\end{align}
The last equation implies that $\rho_{\bar{g}}(\omega)$ has to be both positive and negative for some values of the frequency. A more complete and comprehensive discussion about the graviton spectral functions properties can be found in~\cite{Bonanno:2021squ}.

Once the fluctuation spectral function $\rho_h(\omega)$ has been derived from the Euclidean propagator by using~\cref{eq:ImGE}, we can compute the real-time Minkowskian counterpart by using the following relation
\begin{align}\label{eq:GravitonParam}
	\mathcal{G}_{hh} (p^2) & = \frac{1}{Z_{h} (p)} \frac{1}{p^2+i \epsilon } \notag\\
	& = \frac{1}{p^2 +i \epsilon } + \int_{0}^{\infty}\!\!\mathrm{d} q^2 \frac{\rho_{h}^{\text{cont}}(q^2)}{p^2 - q^2 +i \epsilon} \, ,
\end{align}
where we have separated the contributions from the pole and continuum. The pole provides us with the $1/p^2$ term and the continuum $\rho_h^{\text{cont}}$ gives subleading contributions that become relevant only from the Planck scale onwards, namely when the quantum effects induced by gravity cannot be neglected anymore.

\subsection{Approximation beyond leading order}
\label{sec:our-approximation}

In this section, we explain how we go beyond the leading-order contributions to the cross-section by employing the introduced real-time tools in \Cref{sec:spectral-functions} to upgrade the classical $n$-point functions with their fully momentum-dependent $1\text{PI}$ counterparts from the quantum effective action. This corresponds to the translation
\begin{align}
	\label{eq:n-point-identification}
	S^{(n)}(\mathbf p)\longrightarrow \Gamma^{(n)}(\mathbf p)\,,
\end{align}
in \cref{eq:MEgraviton,eq:currents}. The parameterisation of the $\Gamma^{(n)}$ is given explicitly in \Cref{fig:dressed}. Moreover, the classical external legs of the spinor fields in \cref{eq:currents} are also properly upgraded with their renormalised version by including the respective field dressing. This renders the matrix element for the graviton-mediated diagram
\begin{align}\label{eq:upgradeM}
	{\cal M}_h \propto s \, G^\frac12_{\text{N},h\bar\psi\psi}(p_{e^+},p_{e^-},p_h)\, G^\frac12_{\text{N},h\bar\psi\psi}(p_{\mu^+},p_{\mu^-},p_h),
\end{align}
where $p_h^2 = (p_{e^+}+p_{e^-})^2=s$, and $G_{\text{N},h\bar\psi\psi}$ is the Newton coupling avatar from the fermion-graviton vertex defined in \cref{eq:fermion-graviton-newton}. For simplicity, we are only displaying the scalar part of the matrix element and do not show the tensor structures. In \cref{eq:upgradeM} the wave function renormalisations of all fields exactly cancel out, see also \cref{eq:def-gamma-bar}.

In the IR with $G_{\text{N},h\bar\psi\psi} (\mathbf p \to 0) = G_\text{N}$,  \cref{eq:upgradeM} falls back to the leading-order expression given in \cref{eq:Mh-leading}. Nonetheless, \cref{eq:upgradeM} is still a general expression for the matrix element and is able to capture the full quantum behaviour. The fully momentum-dependent Newton coupling $G_{\text{N},h\bar\psi\psi}$ fulfils a highly complicated integral-differential flow equation that is difficult to solve for general momentum dependencies. In the past, these have been computed in Euclidean signature at the momentum-symmetric point where all momenta have the same magnitude and the scalar products are given by $p_i \cdot p_j = p^2 (3/2 \delta_{ij}-1/2)$ in case of a three-point function. This approximation significantly simplifies the flow equation for the correlation function since only a single momentum variable dependence is retained. In many cases, this is often a decent approximation since the vertices display a mild dependence on the angles between the external momenta, with the exception of exceptional momentum configurations where one of the momenta is vanishing. In summary, we make the replacement
\begin{align}
	G_{\text{N},h\bar\psi\psi}(\mathbf p) \longrightarrow G_{\text{N},h\bar\psi\psi}(p^2) \,.
\end{align}
As discussed in \Cref{sec:eff-act}, $G_{\text{N},h\bar\psi\psi}$ is called an avatar of the Newton coupling, which is in this case extracted from the fermion-graviton vertex. The differences between different avatars of the gravitational coupling, either for pure graviton or graviton-matter couplings, are measured by modified Slavnov–Taylor identities. In~\cite{Eichhorn:2018akn, Eichhorn:2018ydy}, it has been shown that these avatars show a surprisingly strong similarity at the UV fixed point, which was called effective universality.

\begin{figure*}[tbp]
	\includegraphics[width=.48\linewidth]{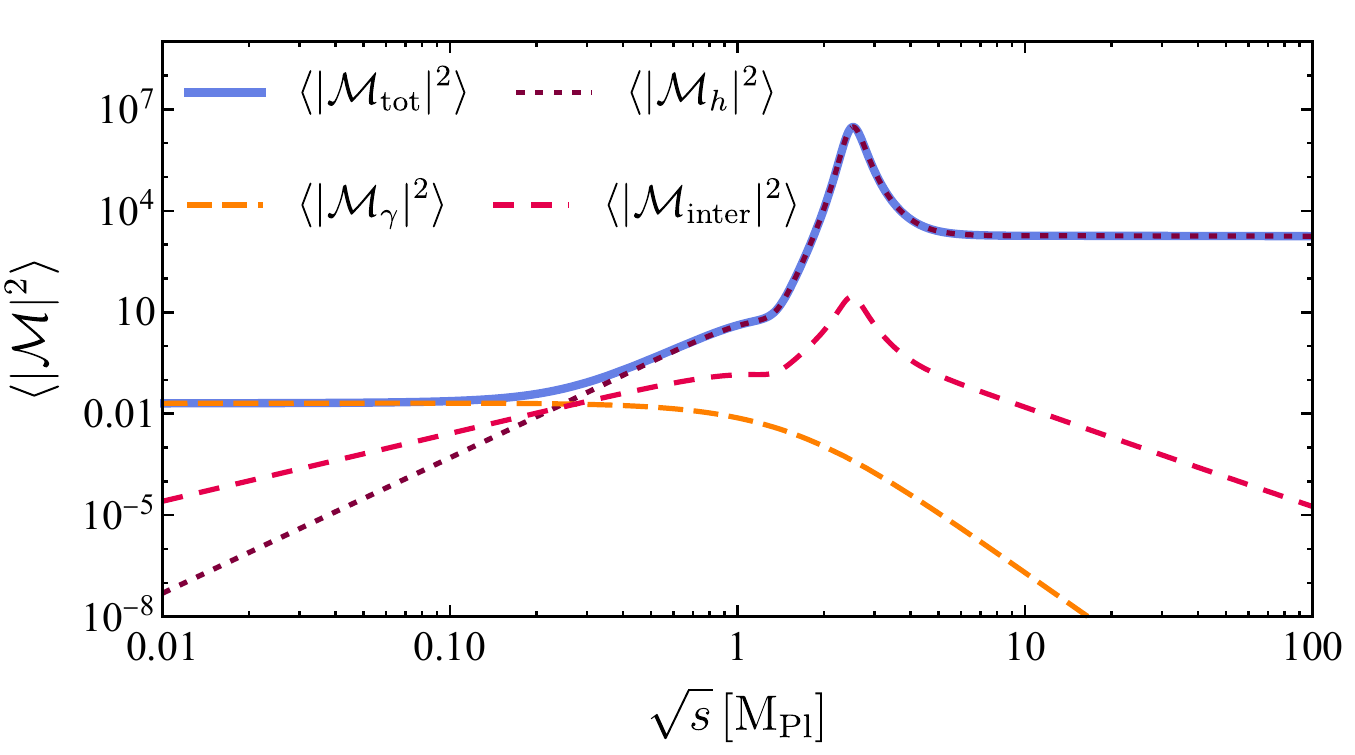} 
	\hfill
	\includegraphics[width=.48\linewidth]{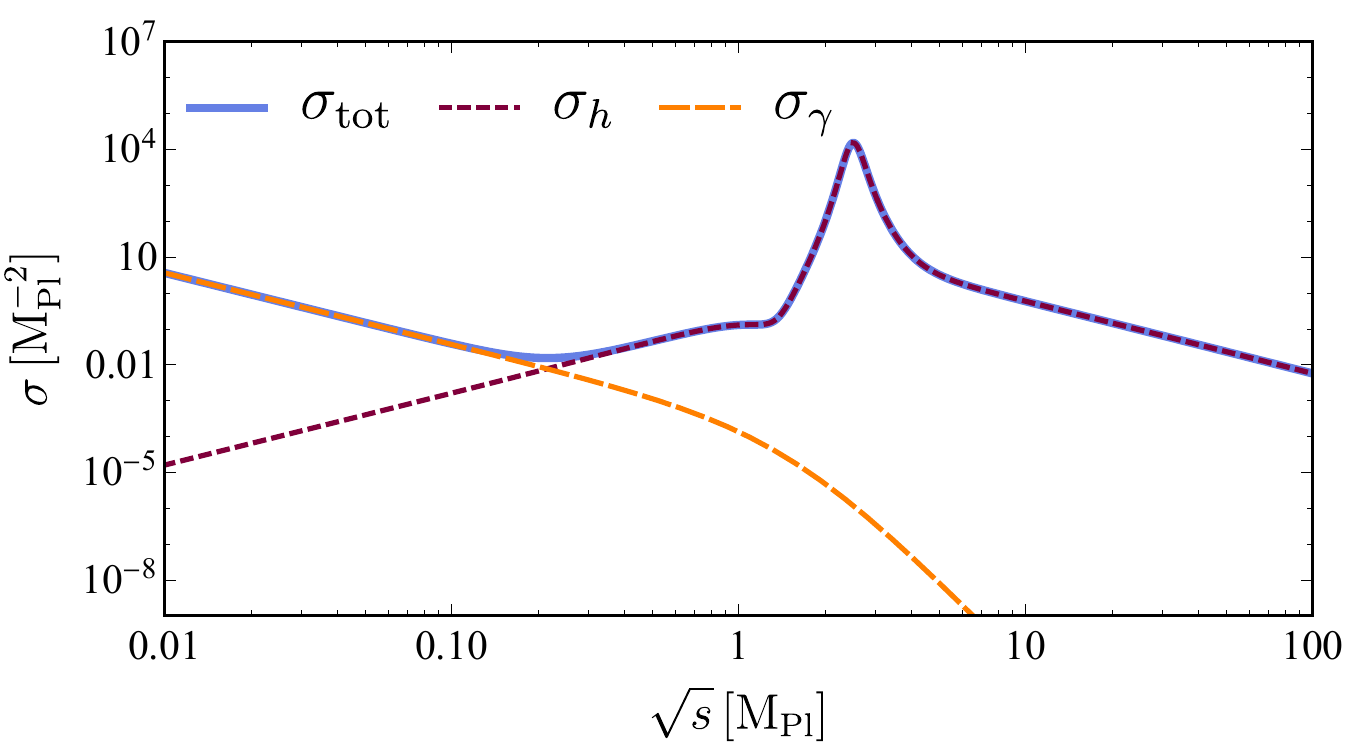} 
	\caption{Left panel: total scattering amplitude \labelcref{eq:TotalM} for $e^+ e^- \rightarrow \mu^+ \mu^- $ (solid blue line) at scattering angle $\theta = \pi/4$, and the contributions of the different diagrams. The dashed orange and dotted purple lines depict the photon-mediated and graviton-mediated contributions. The dashed red line shows the contribution provided by the interference between the photon-mediated and graviton-mediated scattering, i.e.~$ \langle \vert \mathcal{M}_{\rm inter} \vert^2 \rangle \equiv \langle \vert \mathcal{M}_{\gamma}^\ast \mathcal{M}_h + \mathcal{M}_{h}^\ast \mathcal{M}_{\gamma} \vert \rangle$. Right panel: total cross section \labelcref{eq:sigmaTOTbeyondLO} for $e^+ e^- \rightarrow \mu^+ \mu^- $ (solid blue line), and the contributions of the different diagrams. The dashed orange and dotted purple lines depict the photon-mediated and graviton-mediated contributions.} 
	\label{fig:sigmatot}
\end{figure*}

The avatar of the Newton coupling from the TT three-graviton vertex is the one that has been predominantly studied in the literature \cite{Christiansen:2015rva, Denz:2016qks}. The momentum dependence at vanishing RG scale has been studied and the analytic continuation to Lorentzian signature has been performed \cite{Bonanno:2021squ}. Furthermore, we expect that the three-graviton vertex is best suited to capture the intricate momentum dependencies of quantum gravity fluctuations. In this work, we build on the results from the three-graviton vertex and identify the Newton coupling from the fermion-graviton vertex with that of the three-graviton vertex,
\begin{align}
	G_{\text{N},h\bar\psi\psi}(p^2) \longrightarrow  	G_{\text{N},3h}(p^2) \,.
\end{align}
For the computation of graviton-mediated scattering, we need this coupling in Lorentzian signature for timelike momenta. In \cite{Bonanno:2021squ}, the Newton coupling $G_{\text{N},3h_{tt}}$ has been accurately computed in Euclidean signature, and then subsequently continued analytically to Lorentzian signature,
\begin{align}\label{eq:analytic-continuation}
	G_{\text{N},3h}(p_E^2) \xrightarrow[\text{continuation}]{\text{analytic}} 	G_{\text{N},3h}(p^2) \,.
\end{align}
In \cite{Bonanno:2021squ}, this analytic continuation was achieved by utilising the background propagator $\mathcal{G}_{\bar{g} \bar{g}}$ with its spectral representation $\rho_{\bar{g}}$  given in \Cref{fig:spectral-function}. The background spectral function is directly related to the physical momentum-dependent $G_{\text{N},3h}(p^2) $ extracted from the graviton three-point function. This relation was derived in \cite{Bonanno:2021squ} by contracting the external legs with two further fluctuation graviton propagators,
\begin{align}\label{eq:BackGN}
	\mathcal{G}_{\bar{g}\bar{g}}(p^2) &\propto \hspace{0cm}\mathcal{G}_{hh}(p^2)  [\Gamma^{(hhh)}(p^2) \, \mathcal{G}_{hh}(p^2) \, \Gamma^{(hhh)}(p^2)]  \mathcal{G}_{hh}(p^2)\notag\\[2ex]
	&\propto\vcenter{\hbox{\includegraphics{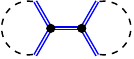}}}\,\,.
\end{align}
From here, all fluctuation wave-function renormalisations drop out and we obtain the following relation
\begin{align}\label{eq:backprop-threegravcoupling}
	\mathcal{G}_{\bar{g}\bar{g}}(p)=\frac1{ Z_{\bar g} (p^2)\,  p^2} \propto \frac{G_{\text{N},3h}(p^2)}{p^2}\,.
\end{align}
We remark that we are using exactly the results derived in \cite{Bonanno:2021squ}, which implies that fluctuation propagator $\mathcal{G}_{hh}$ and the background propagator $\mathcal{G}_{\bar{g}\bar{g}}$ were computed in an approximation where the cosmological constant was set to zero and the impact of matter fluctuations was neglected. The running of the SM couplings is computed in straight analogy with \cite{Pastor-Gutierrez:2022nki} but with this approximation of the Newton coupling.

In conclusion, with the given approximations, the matrix element for the graviton-mediated process reads
\begin{align}
	\label{eq:final-Mh}
	{\cal M}_h \propto  s\, G_{\text{N},3h}(s) \propto \frac{s}{Z_{\bar g} (s)} \,,
\end{align}
which is sensitive to the real-time features of the graviton propagator. 

We now upgrade the photon-mediated contribution given in \cref{eq:Mgamma-leading} with quantum gravity effects by replacing the classical correlation functions to the full quantum analogues, see \cref{eq:n-point-identification}. The photon-mediated process is dominant at low energies but becomes subdominant compared to the graviton-mediated process at the Planck scale. This is evident since the leading-order process scales with $1/s$, see \cref{eq:Mgamma-leading}. Therefore, we do implement a less accurate approximation for the photon-mediated matrix element. Specifically, we resort to an RG-improvement of the coupling which means we identify the RG-scale dependence with the centre-of-momentum energy. The coupling in \cref{eq:fermion-photon} is replaced with
\begin{align}\label{eq:RGI-electriccoupling}
	\alpha_e(k, \mathbf p)\longrightarrow \alpha_{e}(k = s)\,.
\end{align}
With this approximation, we do include explicit real-time momentum dependencies in the photon-fermion vertices but account for the secondary source of gravitational effects entering through the graviton corrections to these couplings. The RG dependence of the coupling is computed in the asymptotically safe Standard Model \cite{Pastor-Gutierrez:2022nki} and becomes asymptotically free beyond the Planck scale.

\section{$e^+e^- \rightarrow \mu^+\mu^-$ in the Asymptotically Safe Standard Model}
\label{sec:scattering-ASSM} 

In this section, we discuss the total $e^+ e^- \rightarrow \mu^+ \mu^-$ cross section derived in \Cref{sec:cross-section} and improved beyond leading order employing the real-time tools introduced in \Cref{sec:real-time-quantum-eff-action,sec:our-approximation} in the gravity sector and RG-improving the matter sector. We also compare our results to other approximation schemes such as a next-to-leading computation and performing different types of RG-improvement in the graviton-meditated cross-section.

\begin{figure*}[ht!]
	\centering
	\includegraphics[width =0.625\columnwidth]{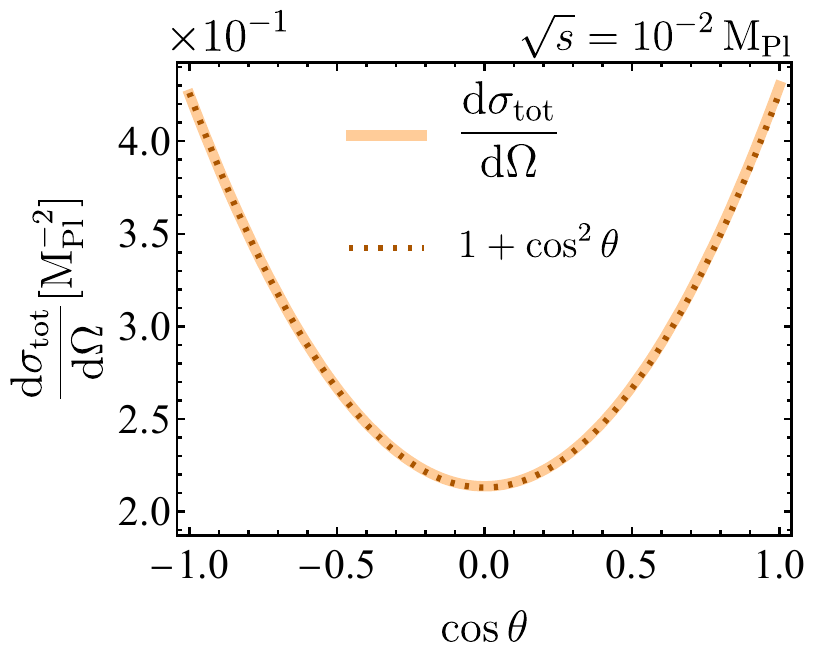} \hspace{.5cm}
	\includegraphics[width =0.625\columnwidth]{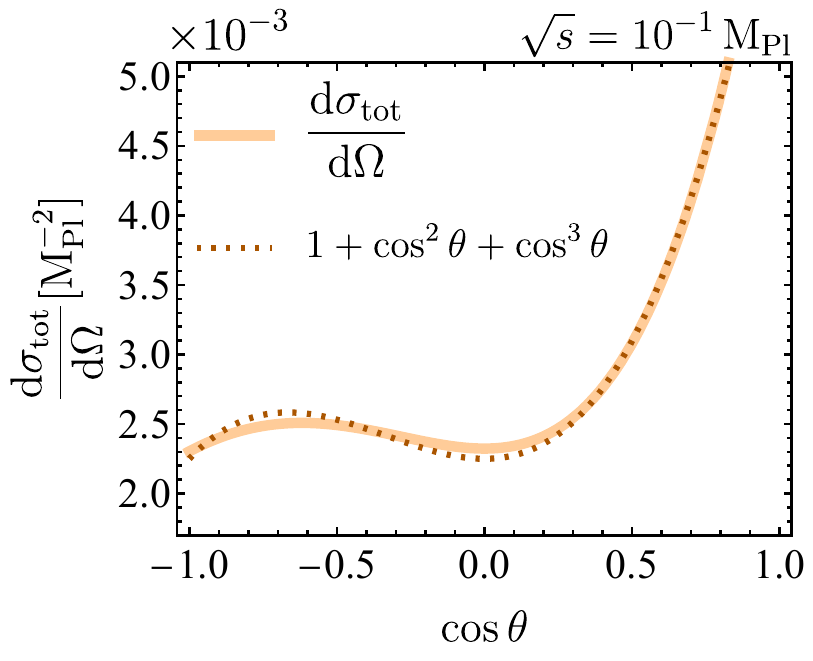}\hspace{.5cm} 
	\includegraphics[width =0.625\columnwidth]{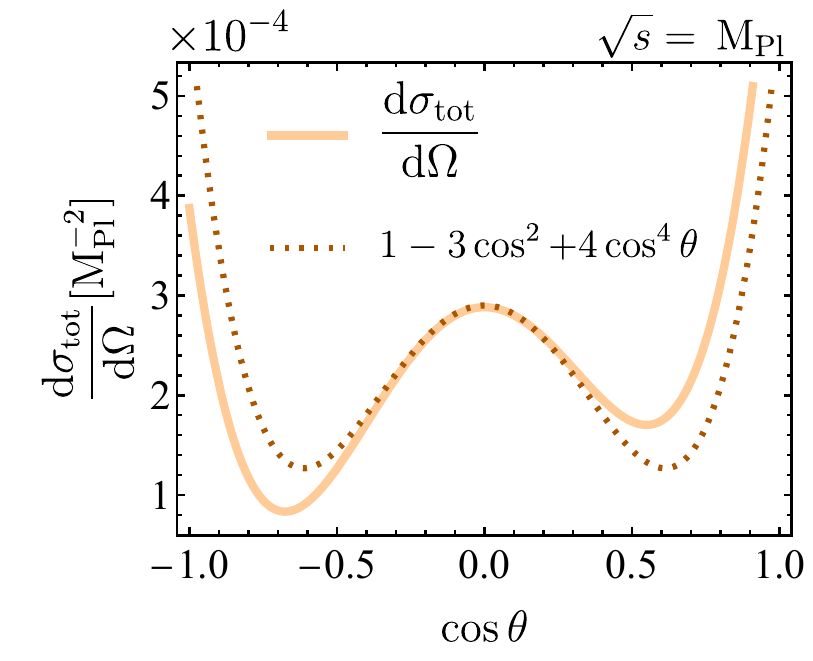} 
	\caption{Total differential cross-section as a function of the angular dependence for three different centre-of-mass energies. For $\sqrt{s}<\text{M}_{\text{Pl}}$, the $(1 + \cos^2 \theta)$ dependence dominates. For $\sqrt{s} \approx 10^{-1} \text{M}_{\text{Pl}}$, an additional contribution with $\cos^3 \theta$ provided by the interference terms $\mathcal{M}_{\gamma}^{\ast} \mathcal{M}_{h}$ and $\mathcal{M}_{\gamma} \mathcal{M}_{h}^{\ast}$ appears and becomes relevant. From the Planck scale onwards, the angular dependence is given by the $(1 - 3 \cos^2 \theta + 4 \cos^4 \theta)$ term provided in $\vert \mathcal{M}_{h} \vert^2$.}
	\label{fig:dsigmaTOT}
\end{figure*}

\subsection{Non-perturbative cross-section}
\label{sec:full-cross-sections}
Summing up the amplitudes shown in~\Cref{sec:matrix-element} and accounting for quantum gravity effects, the total non-perturbative amplitude is given by
\begin{align}
\langle \vert \mathcal{M}_{\rm tot} \vert^2 \rangle  = &~ 16 \pi^2 \alpha_e(s)^2 (1 + \cos^2 \theta) + \alpha_e(s) \frac{s}{Z_{\bar g} (s)} \cos^3 \theta \notag  \\
& + \pi^2 \frac{s^2} {Z_{\bar g}^2 (s)} ( 1 - 3 \cos^2 \theta + 4 \cos^4 \theta )\,.
\label{eq:TotalM} 
\end{align}
The scattering amplitude is a dimensionless quantity and in the scale invariant trans-Planckian regime it approaches a constant value governed by the interacting UV fixed point. This is depicted at fixed scattering angle $\theta$ by the solid blue line in the left panel of~\Cref{fig:sigmatot}. The other lines display the partial contributions to the total amplitude. 
Compared to the full amplitude and the graviton-mediated contribution, the photon-mediated contribution decays in the trans-Planckian regime as this contribution is governed by the Gau\ss ian fixed point of the electromagnetic coupling.

Similarly, collecting the results from previous sections, we arrive at the full non-perturbative cross-section,
\begin{align}
	\sigma_{\rm tot }(s) = \frac{4 \pi \alpha_{e}(s)^2}{3 s} +  \frac{\pi}{20}\frac{s}{Z^2_{\bar g} (s) } \,,
\label{eq:sigmaTOTbeyondLO}
\end{align}
which contains the quantum effects of gravity and real-time features of the graviton propagator.
In the right panel of~\Cref{fig:sigmatot}, we show the total cross-section and partial contributions as a function of the centre-of-mass energy. 

The graviton contribution is significantly subdominant in the IR and only at  $\sqrt{s}\gtrsim {\rm M}_{\rm Pl}$ overtakes the photon-mediated one. A striking feature of this cross-section is a prominent peak appearing at $\sqrt{s}\sim 2\cdot 10^{19}$\,GeV. This peak comes technically from a peak in the background graviton spectral function depicted in the right panel of \Cref{fig:spectral-function}. Therefore this is a real-time feature and may carry physical information. The peak could be related to resonances of graviton bound states with a mass of the order of the Planck scale, i.e., some temporary formation of quantum black holes, see \cite{Dvali:2014ila, Giddings:2007qq, Dvali:2012en} for a discussion of black hole formation from scattering processes.  Alike features are present in the background spectral function of the gluon propagator in QCD \cite{Cyrol:2018xeq}. Nonetheless, we want to caution that this might also be an artefact of the present approximation. 

The photon-mediated scattering dominates in the IR. At the Planck scale, the slope of the cross-section changes. This is caused by the running of the electroweak and hypercharge gauge couplings which evolve logarithmically below the Planck scale and are attracted towards the Gau\ss ian fixed point with a power law scaling above the Planck scale, see \cite{Pastor-Gutierrez:2022nki} for details.

Most importantly, for $\sqrt{s}\gg M_\text{Pl}$, the total cross section decreases with $1/s$. This is caused by the momentum dependence of the background graviton propagator which scales with $1/p^4$ for large momenta, or equivalently the momentum dependence of the Newton coupling, which scales with $1/p^2$. Both behaviours are equivalent and are caused by fixed-point scaling. 

At leading-order, the cross-section violates unitarity as it increases with $s$. With the full non-perturbative quantum corrections included, the cross-section decreases with $1/s$. Therefore it obeys the Froissart bound \cite{Froissart:1961ux} and is compatible with unitarity.

The interference terms $\mathcal{M}_{\gamma}^{\ast} \mathcal{M}_{h}$ and $\mathcal{M}_{\gamma} \mathcal{M}_{h}^{\ast}$ do not contribute to the total cross-section $\sigma_{\rm tot}$, as they vanish upon integration over the full angular dependence. However, the differential cross-section~\cref{eq:dsigmaTOT} still exhibits non-trivial dependence, as shown in~\cref{fig:dsigmaTOT} for different values of the centre-of-mass energy $\sqrt{s}$ and angular configurations. For this analysis, we improved the differential cross-section following the arguments in \Cref{sec:our-approximation}, where only symmetric-point momentum dependencies are considered. We emphasize that in this case, the full momentum dependence on different combinations of the external leg momenta can lead to non-trivial features, which are included in the fully momentum-dependent Newton coupling $G_{\text{N},h\bar\psi\psi}(\mathbf p)$ but whose effects are not accounted for here.

The angular dependence of the differential cross-section is an important physical observable that can be used to determine whether gravity-fermion systems exhibit parity-conserving interactions. In our specific case, we observe a forward-backward symmetric distribution when considering only gravitational interactions, implying that equal numbers of muons would be produced in the forward hemisphere $(\cos \theta > 0)$ as in the backward hemisphere $(\cos \theta < 0)$. In the case of the asymptotically safe SM, this symmetry would be realised from the Planck scale onward, as all other interactions and their interferences with gravity can be neglected.

\subsection{Comparison to other approximations}
\label{sec:comparions-beyondLO}
Here we compare the full non-perturbative result to other approximations. First, we do a perturbative next-to-leading order (NLO) approximation and secondly, we compare our result to using RG improvement in the gravitational-mediated cross-section.

\subsubsection{Next-to-leading order}
\label{sec:NLO}
Performing a perturbative expansion of the non-perturbative graviton-mediated cross-section, the expanded cross-section reads
\begin{align}\label{eq:sigma-eff-NLO}
	\sigma_h(s) = \sigma_h^{\rm LO} (s)+ \sigma_h^{\rm NLO} (s)+  \mathcal{O}(G_\text{N}^5)\, ,
\end{align}
where $	\sigma_h^{\rm LO} $ is the leading graviton contribution given in \cref{eq:sigmaTOT}. The contribution $\sigma_h^{\rm LO}$ is purely tree-level and is proportional to $G_\text{N}^2$. The contribution contains mixed terms between the tree-level and one-loop amplitude, $\mathcal M_{h,\text{tree}}^* \mathcal M_{h,\text{1-loop}}$, that are proportional to $G_\text{N}^3$, and also the full one-loop contribution, $\mathcal{M}_{h,\text{1-loop}}^* \mathcal{M}_{h,\text{1-loop}}$, that is proportional to $G_\text{N}^4$. 

The NLO contribution is obtained by expanding the background propagator for small centre-of-mass energies $s \ll {\rm M}_{\text{Pl}}$, see \cref{eq:final-Mh}. The LO behaviour of the propagator is the classical $1/s$. At NLO, we find a logarithmic behaviour as it is typically arising from the universal logarithmic divergences of the theory. Conceptually, this is similar to the universal logarithms such as $R \log(\Box) R$ appearing in the one-loop effective action of quantum gravity. The resulting NLO part of the graviton-mediated cross-section reads
\begin{align}\label{eq:NLO}
	\sigma_h^{\rm NLO}(s) & = \frac{\pi}{5}  A_{\bar{g}} s^2 G_{\rm N}^3 \left[ \gamma_{\rm E} + \ln( s G_{\rm N}) \right] \\[1ex]
	&\hspace{-.2cm}+ \frac{\pi}{5}  A_{\bar{g}}^2 s^3 G_{\rm N}^4 \left[ 1 + \gamma_{\rm E}^2 + 2 \gamma_{\rm E}\ln(s G_{\rm N}) + \ln^2( s G_{\rm N}) \right].\notag
\end{align}
Here, $\gamma_{\rm E}$ is the Euler-Mascheroni constant and $A_{\bar{g}} = -111/(380 \pi)$. The coefficient $A_{\bar{g}}$ is precisely the coefficient of the logarithmic term of the background propagator determined in \cite{Bonanno:2021squ}. The coefficient is regulator-independent but depends on the gauge-fixing parameters. Note that the Euler-Mascheroni constant $\gamma_{\rm E}$ should not appear in a measurable scattering amplitude. This is an artefact from our approximation, e.g., from neglecting NLO terms coming from the third diagram in \Cref{fig:scattering-diagrams}. 

As expected, the NLO contribution in \cref{eq:NLO} does not respect the Froissart bound and violates unitarity. The cross-section diverges even faster than at leading order. This highlights that indeed a full non-perturbative result is necessary to analyse the unitarity of a given cross-section. 

\subsubsection{RG improvement}
\label{sec:RGi}
RG improvement is an often-used method to obtain a qualitative estimate of the scaling of a given quantity. It typically works well in systems with only one physical scale. The cross-section depends on the Mandelstam variables $s$, $t$, and $u$ where the latter two can be expressed in terms of $s$ and the scattering angle $\theta$. Therefore it is natural to identify the RG scale with the centre-of-mass energy, $k\to s$.  In this subsection, we check how well the identification works in comparison with our full result. As a second approximation, we compare our result to the case where we use the full Euclidean momentum dependence of the Newton coupling without the analytic continuation given in \cref{eq:analytic-continuation}.

\begin{figure}[tbp]
	\includegraphics[width=\columnwidth]{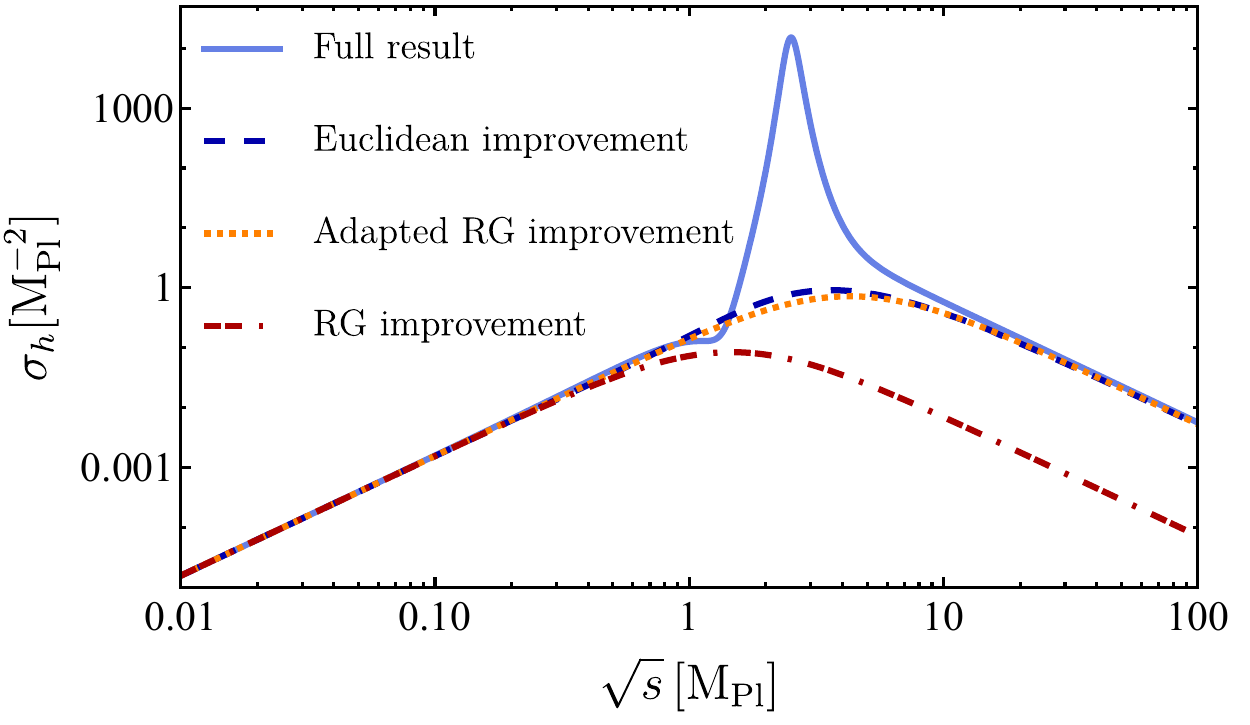} 
	\caption{Graviton-mediated contribution to the $e^+ e^- \to \mu^+ \mu^-$ cross-section in different approximations. The solid blue line depicts the full result, see \cref{fig:sigmatot}. The dashed dark blue line represents the cross-section after an RG improvement with the Euclidean momentum-dependent Newton coupling \cref{eq:Euclidean-improvement}. The red dash-dotted line shows the RG improved cross-section where the cutoff is identified with the centre-of-mass energy, see \cref{eq:cutoff-improvement}, and the dotted orange line shows the RG improved cross-sections with the parametrised Newton coupling in \cref{eq:GN-parametrisation}, employing the momentum fixed-point value in \cref{eq:FP-values2}, which we call adapted RG improvement.}
	\label{fig:gravitycrossections}
\end{figure}

For the purpose of this section, we write the Newton coupling as a function of the Euclidean momentum $p^2$ and the RG scale $k$,
\begin{align}
	G_\text{N}(k,p^2)\,.
\end{align}
There are two suggestive improvements:
\begin{itemize}
	\item[(i)] Using the Newton coupling at vanishing momentum as a function of $k$ and identifying
	\begin{align}\label{eq:cutoff-improvement}
		G_\text{N}(k,0) \longrightarrow 	G_\text{N}(\sqrt{s},0)\,.
	\end{align}
	This is the standard RG improvement and it is depicted by the red dash-dotted line in \Cref{fig:gravitycrossections}.
	\item[(ii)] Using the Euclidean momentum dependence of the Newton coupling at vanishing RG scale and identifying
	\begin{align}\label{eq:Euclidean-improvement}
		G_\text{N}(0,p^2) \longrightarrow 	G_\text{N}(0, s)\,.
	\end{align}
	This is depicted by the dashed dark blue line in \Cref{fig:gravitycrossections} and we call it Euclidean improvement.
\end{itemize}
Both RG improvements provide the correct UV scaling of the cross-section, $\sigma \sim 1/s$, but both approximations miss the peak of the cross-section around the Planck scale since this feature is linked to the timelike momenta of the process. As expected, approximation (ii) correctly estimates the value of the cross-section for large $s$, while approximation (i) underestimates the value by several orders of magnitude. This is linked to the fact that the fixed-point value of the Newton coupling at $p^2=0$ is significantly smaller than the value of the Newton coupling at large momentum and vanishing RG scale,
\begin{align}
	G_\text{N}(k\to\infty,0)  &= 2.15 \,, \notag \\
	G_\text{N}(0,p^2 \to \infty)  &= 18.4 \,,	\label{eq:FP-values2}
\end{align}
see \cite{Bonanno:2021squ} for more details.

The simple shapes of the RG improved cross sections and the difference in the RG and momentum fixed-point values suggests a third way of using RG improvement:
\begin{itemize}
	\item[(iii)] Use a simple trajectory of the Newton coupling with the momentum fixed-point value and identify the RG scale with the centre-of-mass energy $s$,
	\begin{align}\label{eq:GN-parametrisation}
		G_\text{N}(s) = \frac{g^\ast}{s + g^\ast \, {\rm M}_\text{Pl}^2},
	\end{align}
	where $g^\ast = 	G_\text{N}(0,p^2 \to \infty) $.	We call this the adapted RG improvement and it is depicted as the orange dotted line in \Cref{fig:gravitycrossections}.
\end{itemize}
This approximation captures the UV asymptotics correctly and only requires the computation of the momentum fixed-point value, $G_\text{N}(0,p^2 \to \infty)$.

\section{Conclusions}
\label{sec:conclusion}

We have studied the $e^+e^- \rightarrow \mu^+\mu^-$ scattering process in asymptotically safe quantum gravity. First, we computed the leading order photon and graviton contributions to the cross-sections. As expected the leading order violates unitarity bounds as the cross-section increases with the centre-of-mass energy. Subsequently, we computed the non-perturbative cross-section by employing 1PI correlation functions from the quantum effective action. This cross-section decays with the centre-of-mass energy beyond the Planck scale and therefore is compatible with unitarity requirements. Our work therefore presents significant evidence in favour of the unitarity of asymptotically safe quantum gravity.

Our work is the first to take into account non-perturbative real-time effects in the graviton-mediated cross-section. This was achieved through the spectral representation of the graviton propagator, which bridges Euclidean results to Minkowski space. Including such contributions leads to the appearance of a peak structure at centre-of-mass energies close to the Planck scale. Although this feature resembles a resonance, within the present approximation, further work has to be invested for confirming its physical significance.

Furthermore, we have compared the present results to other approximations, such as RG improvement. While all approaches show similar asymptotic scaling, the spacelike momentum couplings lack the peak feature. Additionally, we showed that the adapted RG improvement employing the fixed-point coupling in \cref{eq:FP-values2} provides a simple and efficient approximation to the full result.

In the current computation, several approximations were made that will be addressed in future work. Most notably, we approximated the fermion-graviton coupling with that of the three-graviton vertex. In future studies, we want to include the explicit real-time momentum dependence of the former relevant coupling. Moreover, we neglected the four-fermion contact terms in \Cref{fig:scattering-diagrams}. We hope to report on the respective improvements as well as applications of the present approach to a comprehensive set of cross sections in the near future.

\section*{Acknowledgements}
We thank T.~Denz and M.~Yamada for discussions and T. Denz for technical support on V{\footnotesize ERT}EX{\footnotesize PAND}. M.R. acknowledges support by the Science and Technology Facilities Council under the Consolidated Grant No. ST/X000796/1 and the Ernest Rutherford Fellowship  No. ST/Z510282/1. This work is funded by the Deutsche Forschungsgemeinschaft (DFG, German Research Foundation) under Germany’s Excellence Strategy EXC 2181/1 - 390900948 (the Heidelberg STRUCTURES Excellence Cluster) and the Collaborative Research Centre SFB 1225 (ISOQUANT).

\appendix

\section{Theoretical framework}

\subsection{Classical action}
\label{app:classical-action}
The classical Einstein-Hilbert action is given by 
\begin{align}
	S_{\text{EH}}[g_{\mu \nu}] = \frac{1}{16 \pi G_\text{N}} \int \! \mathrm{d}^{4}x \, \sqrt{g} \, \Big( 2 \Lambda - R(g_{\mu \nu}) \Big),
\end{align}
with the classical Newton constant $G_{\text{N}}$ and the abbreviation $\sqrt{g} = \sqrt{|\det g_{\mu \nu}(x)|}$. The definition of a graviton propagator requires gauge fixing. In turn, a standard linear gauge fixing requires the definition of a background metric, which also serves as the expansion point of the effective action. Here, we implement a linear split for the full metric $g_{\mu \nu} = \bar g_{\mu \nu} +\sqrt{G_\text{N}} h_{\mu  \nu}$, with the background metric given by the flat Minkowski metric $\bar g_{\mu\nu}= \eta_{\mu\nu} = \text{diag}(+1, \mathbf{-1})$. 
In this way, the gauge-fixing action reads
\begin{align}
	S_{\text{gf}}[h] = \frac{1}{32 \pi \alpha} \int \! \mathrm{d}^{4}x  \, F_\mu F^\nu  ,
\end{align}
where a De-Donder-type linear gauge with $\alpha \to 0$ has been used
\begin{align}
	F_\mu = \bar{\nabla}^\nu h_{\mu \nu} - \frac{1 + \beta}{4} \bar{\nabla}_\mu h^{\nu}_{\, \, \nu} \, ,
\end{align}
and $\bar{\nabla}$ indicates the ordinary covariant derivative in the flat Minkowski space.
The ghost action corresponding to the gauge-fixing condition is given by
\begin{align}
	S_{\text{gh}}[h, \bar{c}, c] & = \int \! \mathrm d^4x \,  \bar{c}^\mu \, \mathcal{M}_{\mu \nu} \, c^\nu ,
\end{align}   
with $\mathcal{M}_{\mu \nu}$ being the Faddeev-Popov operator, which can be stemmed from a diffeomorphism transformation of the gauge fixing condition $F_\mu$
\begin{align}
	\mathcal{M}_{\mu \nu} = \bar{\nabla}^\rho \left( g_{\mu \nu} \nabla_\rho + g_{\rho \nu} \nabla_{\mu}\right) - \frac{1 + \beta}{2}\bar{\nabla}_\mu \nabla_\nu \,.
\end{align}
The classical action for minimally-coupled fermions fields is given by 
\begin{align}
	\label{formula:Sint}
	S_{\text{int}}[g, \bar{\psi}, \psi] = \int \mathrm{d}^{4}x \sqrt{g} \, \bar{\psi} \left(  i \slashed{\nabla} - m \right) \psi \, .
\end{align}
For the formulation of fermions in curved spacetime, the spin-base invariance formalism~\cite{Weldon:2000fr, Gies:2013noa, Lippoldt:2015cea} has been implemented
\begin{align}
	\slashed{\nabla} = g_{\mu \nu} \gamma^\mu (x) (D^\nu + \Gamma^\nu (x)) \, ,
\end{align}
where the $\gamma$'s are the spacetime dependent Dirac matrices and $\Gamma^\mu$ is the spin connection. Last, $D^{\nu}$ is the ordinary covariant derivative
\begin{align}
	D^{\nu} = \partial^{\nu} - i e A^{\nu}.
\end{align}

\subsection{Graviton-fermion vertex}
\label{app:grav-ferm-vertex}

The vertex tensor structure of the graviton-fermion-antifermion three-point function reads
\begin{widetext}
	\begin{align}
		\label{formula:vertex}
		\Big[ \mathcal{T}^{(h \psi \bar{\psi})} (p_\psi, p_{\bar{\psi}}) \Big]^{\mu \nu}  = -\frac{i}{8} \, \delta^{(4)}(p_{\psi} + p_{\bar{\psi}} + p_h) \,
		\left\lbrace  4  m_{\psi}  \eta^{\mu \nu} + \gamma^{\nu}(p_\psi^{\mu} -p_{\bar{\psi}}^{\mu}) 
		+ \gamma^{\mu}(p_\psi^{\nu} -  p_{\bar{\psi}}^{\nu}) 
		- 2 \eta^{\mu \nu} (\slashed{p}_\psi -  \slashed{p}_{\bar{\psi}})
		\right\rbrace  \, .
	\end{align}  
\end{widetext}
For the derivation of the n-point functions, we relied on the Mathematica package V{\footnotesize ERT}EX{\footnotesize PAND}~\cite{vertexpand}.
The tensor structure of the vertex is in agreement with~\cite{Coriano:2013iba, Cho:1992hx, Holstein:2006bh}. In~\cref{formula:vertex}, it is implied that the momenta of the external fermions both enter or leave the vertex. 
We show here a fundamental property of this vertex that every other Standard Model vertex satisfies, namely
\begin{align}
	\label{eq:Vertex_property}
	\left[ \bar{\psi} \, \Gamma \,\phi \right]^{\dagger} \equiv \bar{\phi} \, \Gamma \, \psi ,
\end{align}
with $\psi$ and $\phi$ being two arbitrary fermionic fields and $\Gamma$ is the interaction vertex that only accounts for the tensor structure, the quantity in curly brackets in~\cref{formula:vertex}.
Starting from the last equation
\begin{align}
	\left[ \bar{\psi} \, \Gamma \, \phi \right]^\dagger = \left[ \psi^\dagger \gamma_0 \, \Gamma \, \phi \right]^\dagger = \phi^\dagger \Gamma^\dagger \gamma_0^\dagger \psi \, ,
\end{align}
with $\gamma_0$ being the timelike Dirac gamma matrix.
By inserting the identity matrix between $\phi^\dagger$ and $\Gamma^\dagger$, we obtain
\begin{align}
	\phi^\dagger \, \mathbb{1} \, \Gamma^\dagger \gamma_0^\dagger \psi =  \phi^\dagger \gamma_0 \, \gamma_0 \, \Gamma^\dagger \gamma_0^\dagger \psi = \bar{\phi} \, \gamma_0 \, \Gamma^\dagger \gamma_0 \, \psi = \bar{\phi} \,  \bar{\Gamma} \, \psi  .
\end{align}
Here, we have used that $\gamma_0$ is Hermitian and are denoting
\begin{align}
	\bar{\Gamma} \equiv \gamma_0  \,  \Gamma^{\dagger} \, \gamma_0.
\end{align}
Therefore, we need to verify that
\begin{align}
	\bar{\Gamma} \equiv \gamma_0 \, \Gamma^{\dagger} \,\gamma_0 \overset{?}{=} \Gamma \, ,
\end{align}
for~\cref{eq:Vertex_property} to be valid for gravitational interactions as well. This is straightforward when one takes into account that the tensor structures~\cref{formula:vertex} stick to the following properties
\begin{align}
	& \gamma_0 \, (\eta^{\mu \nu})^\dagger \, \gamma_0 = (\gamma_0)^2 \, \eta^{\mu \nu}  =  \eta^{\mu \nu} , \notag \\[1ex]
	& \gamma_0 \left[ \eta^{\mu \nu} \, \slashed{p} \right]^\dagger \gamma_0 = \, \gamma_0 \, \slashed{p}^\dagger (\eta^{\mu \nu})^\dagger \gamma_0 =  \, \gamma_0 \, \gamma_0 \, \slashed{p} \, \gamma_0 \, \eta^{\mu \nu} \,  \gamma_0 =  \, \eta^{\mu \nu} \,\slashed{p}, \notag \\[1ex]
	& \gamma_0 \left[ \gamma^{\mu} \,  p^{\nu} \right]^\dagger \gamma_0 = \gamma_0 \left[ p^\nu \, (\gamma^{\mu})^\dagger \right] \gamma = \gamma_0 \, p^\nu \, \gamma_0 \, \gamma^\mu \, \gamma_0 \, \gamma_0 = \gamma^{\mu} \, p^\nu \, ,
\end{align}
where we have implemented the following identity $(\gamma^\mu)^\dagger = \gamma_0 \gamma^\mu \gamma_0$ for the derivation of the last two equations. Once this property has been demonstrated, one can easily compute $-i \mathcal{M}_{fi}^\ast$, namely the complex conjugate of the matrix element. 

\subsection{Graviton propagator}
\label{app:propagator}

The fluctuation field $h_{\mu \nu}$ can be decomposed in terms of the transverse-traceless tensor mode $h^{\text{TT}}_{\mu \nu}$, a vector mode $\xi_\mu$, and two scalar modes. An example of this is the York decomposition~\cite{York:1973ia} where the scalar modes are denoted by $\sigma$ and the trace mode by $h = h_{\mu}^{\mu}$. In the case of a Minkowskian background $\eta$, the York projection operators are given in an arbitrary $d$ dimension in terms of the transversal and longitudinal operators in momentum space 
\begin{align}
	\Pi^{(\text{TT})}_{\mu \nu \rho \sigma} & = \frac{1}{2} \left( \Pi^{\text{T}}_{\mu \rho} \Pi^{\text{T}}_{\nu \sigma} + \Pi^{\text{T}}_{\mu \sigma} \Pi^{\text{T}}_{\nu \rho} \right) - \frac{1}{d -1} \left( \Pi^{\text{T}}_{\mu \nu} \Pi^{\text{T}}_{\rho \sigma}\right) , \notag \\[1ex]
	\Pi^{(\xi)}_{\mu \nu \rho \sigma} & = \frac{1}{2} \left( \Pi^{\text{T}}_{\mu \rho} \Pi^{\text{L}}_{\nu \sigma} + \Pi^{\text{T}}_{\mu \sigma} \Pi^{\text{L}}_{\nu \rho} +  \Pi^{\text{T}}_{\nu \rho} \Pi^{\text{L}}_{\mu \sigma}  + \Pi^{\text{T}}_{\nu \sigma} \Pi^{\text{L}}_{\mu \rho} \right), \notag\\[1ex]
	\Pi^{(h)}_{\mu \nu \rho \sigma} & = \frac{1}{d} \eta_{\mu \nu} \eta_{\rho \sigma}\notag,\\[1ex]
	\Pi^{(\sigma)}_{\mu \nu \rho \sigma} & = \frac{1}{d-1}  \Pi^{\text{T}}_{\mu \nu} \Pi^{\text{T}}_{\rho \sigma}  + \Pi^{\text{L}}_{\mu \nu} \Pi^{\text{L}}_{\rho \sigma} - \frac{1}{d} \eta_{\mu \nu} \eta_{\rho \sigma}.
\end{align}
and the mixing operators of the spin-0 modes read
\begin{align}
	\Pi^{(h \sigma)}_{\mu \nu \rho \sigma} & = \frac{1}{\sqrt{d-1}} \eta_{\mu \nu} \Pi^{\text{L}}_{\rho \sigma} - \frac{1}{d-1\sqrt{d}} \eta_{\mu \nu} \eta_{\rho \sigma},\notag\\
	\Pi^{(\sigma h)}_{\mu \nu \rho \sigma} & = \frac{1}{\sqrt{d-1}} \Pi^{\text{L}}_{\mu \nu} \eta_{\rho \sigma} - \frac{1}{4\sqrt{d-1}} \eta_{\mu \nu} \eta_{\rho \sigma}, \, 
\end{align}
with the well-known transversal and longitudinal projectors
\begin{align}
	\Pi^{\text{T}}_{\mu \nu} &= \eta_{\mu \nu } - \frac{p_\mu p_\nu}{p^2},
	&
	\Pi^{\text{L}}_{\mu \nu} &= \frac{p_\mu p_\nu}{p^2} \,,
\end{align}
respectively. The York projectors span the space of symmetric rank 4 tensors
\begin{align}
	\Pi^{(\text{TT})} + \Pi^{(\xi)} + \Pi^{(h)} + \Pi^{(\sigma)} = \mathbb{1}  ,
\end{align}
which implies we can decompose the fluctuation graviton $2$-point function in the following way
\begin{align}
	\Gamma^{(2h)}_{\mu \nu \rho \sigma} (p, q)  = \sum_{l= 1}^{6} \Gamma^{(2h)}_{(l)}(p^2) \, \Pi^{(l)}_{\mu \nu \rho \sigma} \, \delta^{(4)}(p + q),  
\end{align}
and from this last object we can obtain the graviton propagator simply by inverting the scalar coefficients \\
$\mathcal{G}_{hh,(l)} = \left(\Gamma^{(2h)}_{(l)} \right)^{-1}$ according to
\begin{align}
	\Big[ \mathcal{G}_{hh} (p,q) \Big]_{\mu \nu \rho \sigma} = \sum_{l = 1}^{6} \mathcal{G}_{hh,(l)}(p^2) \, \Pi^{(l)}_{\mu \nu \rho \sigma} \, \delta^{(4)}(p + q),  
\end{align}
with the sum running over $l   \in \lbrace (\text{TT}),(\xi), (h), (h \sigma), (\sigma h), (\sigma) \rbrace$. We have for the different modes
\begin{align}
	& \mathcal{G}_{hh,(\text{TT})}  = \frac{1}{Z_{h}(p^2)} \frac{32 \, \pi}{ \left( p^2 - 2\, \Lambda\right) }, \notag \\[1ex]
	& \mathcal{G}_{hh,(\xi)}  = \frac{1}{Z_{h}(p^2)} \frac{32 \, \pi \, \alpha}{\left( p^2 - 2 \, \Lambda \, \alpha\right) },  \notag\\[1ex]
	& \mathcal{G}_{hh,(h)}  = \frac{1}{Z_{h}(p^2)} \frac{64 \, \pi \left[p^2\left( \alpha -3\right) + 4 \, \Lambda \, \alpha \right]  }{C(p^2; \Lambda; \alpha, \beta)}, \notag \\[1ex]
	& \mathcal{G}_{hh,(h \sigma)}  = G_{hh,(\sigma h)} = \frac{1}{Z_{h}(p^2)} \frac{64 \,\sqrt{3} \,p^2 \,\pi\, (\alpha - \beta)}{C(p^2; \Lambda; \alpha, \beta)}, \notag \\[1ex]
	& \mathcal{G}_{hh,(\sigma)}  = \frac{1}{Z_{h}(p^2)} \frac{64 \, \pi \left[ p^2 \left(3\alpha - \beta^2 \right) - 4 \, \alpha \, \Lambda \right]}{C(p^2; \Lambda; \alpha, \beta)} \, .
\end{align}
with
\begin{align}
	C(p^2; \Lambda; \alpha, \beta) = p^4 (\beta - 3)^2 - 4 p^2 (3 + 2\alpha - \beta^2) \Lambda + 16 \alpha \Lambda^2 \, .
\end{align}
In this basis only the transverse-traceless mode does not carry \textit{a priori} gauge dependence.

\section{Matrix element and cross section}
\label{app:matrix-element}
Since we are interested in scattering experiments, the relevant states are the momentum eigenstates at $t = \pm \infty$. The latter are generated by the creation operators $a^{\dagger}_p$ at asymptotically early or late times, denoted by $\vert i \rangle$ and $\vert f \rangle$, respectively. The projection of one on the other gives the elements of the scattering, namely the S-matrix elements $S_{fi} = \langle f \vert S \vert i \rangle$. In this work, the initial and final asymptotic states read
\begin{align}
	\vert i \rangle = \vert p_{e^-} \rangle \vert p_{e^+} \rangle, \quad \langle f \vert = \langle p_{\mu^-} \vert \langle p_{\mu^+} \vert ,
\end{align}
respectively. 

In a free theory, where there are no interactions, the S-matrix is simply the identity matrix~$\mathbb{1}$. When interactions occur, the non-trivial part of the S-matrix is given according to 
\begin{align}
	\label{eq:Non-trivialS}
	\langle f \vert S - \mathbb{1} \vert i \rangle = i (2 \pi)^4 \delta^4 \left(\textstyle \sum p \right) \mathcal{M}_{fi}.
\end{align}
Here, $\delta^4 \left(\textstyle \sum p \right)$ is shorthand for 
$\delta^4 \left(\textstyle \sum p_i^{\mu} - \textstyle \sum p_f^{\mu} \right)$ where $p_i^{\mu}$ are the initial particles’ momenta and $p_f^{\mu}$ are the final particles’ momenta. In~\cref{eq:Non-trivialS}, $\mathcal{M}_{fi} = \langle f \vert \mathcal{M} \vert i \rangle$ can be immediately computed by using the Feynman’s rules in momentum-space. The rules used to derive the matrix element $\mathcal{M}_{fi}$ are given in~\cref{app:grav-ferm-vertex} and~\cref{app:propagator}.

Therefore, the matrix element for the graviton-mediated process can be written down as
\begin{align}\label{eq:explicit-matrix-element}
	i \mathcal{M}_{fi} &=\\[1ex]
	&\hspace{-.5cm}\bar{v}(p_{e^+}) \, \Gamma_{\mu_1 \mu_2}^{(h e \bar{e})} \, u(p_{e^-}) \, \mathcal{G}_{hh}^{\mu_1 \mu_2 \nu_1 \nu_2} \, \bar{u}(p_{\mu^-}) \, \Gamma_{\nu_1 \nu_2}^{(h \mu \bar{\mu})} \, v(p_{\mu^+}) ,\notag
\end{align}
where $ \Gamma^{(h e \bar{e})}$, $\Gamma^{(h \mu \bar{\mu})}$, and  $\mathcal{G}_{hh}$ are the vertices for the graviton-electron-positron, the graviton-muon-antimuon interaction, and the graviton propagator, respectively. The procedure to stem $\mathcal{G}_{hh}$ from $\Gamma^{(2h)}$ is outlined in~\cref{app:propagator} as well. At this point, we are interested in the computation of the matrix element squared, $\mathcal{M}_{fi} = \vert \langle f \vert \mathcal{M} \vert i \rangle \vert^2$, entering into the final result of this work, i.e., the cross-section $\sigma_h$. To this end, it is necessary to find the complex conjugate of~\cref{eq:explicit-matrix-element}. The graviton-fermion vertex fulfils an important property, see~\cref{eq:Vertex_property}, which is also satisfied by every other interaction vertex in the Standard Model and can be written down in the following way
\begin{align}
	\label{eq:VertexProp}
	\left[ \bar{\psi} \Gamma \phi \right]^{\dagger} \equiv \bar{\phi} \Gamma \psi ,
\end{align}
where $\psi$ and $\phi$ are two arbitrary fermionic fields and $\Gamma$ is the interaction vertex containing the tensor structure. By using~\cref{eq:VertexProp}, it is straightforward to derive the complex conjugate of the matrix element from~\cref{eq:explicit-matrix-element}.

After that, we are left with averaging the spin states of the incoming particles and summing over all possible spin states for the outgoing particles
\begin{align}
	\langle \vert \mathcal{M}_{fi} \vert^2 \rangle  = \frac{1}{4} \sum_{\text{spin}} \vert \mathcal{M}_{fi} \vert^2 \, .
\end{align}
It is essential to note that so far we have not yet imposed that asymptotic states are on-shell. This means that the gauge dependence embedded in the propagator is still present. To conclude this section, we will only describe schematically what was done to obtain the physical scattering amplitude. 

First of all, given the large number of tensor contractions, we made use of Form~\cite{Vermaseren:2000nd, Kuipers:2012rf} and Mathematica. Precisely, the F{\footnotesize ORM}T{\footnotesize RACER} package~\cite{Cyrol:2016zqb} was used to trace the diagrams shown in~\cref{fig:scattering-diagrams}. Every scalar product in $\langle \vert \mathcal{M}_{fi} \vert^2 \rangle$ obtained after the tracing can be replaced with other Lorentz-invariant quantities, namely the Mandelstam variables $s, t$, and $u$. 

\begin{figure}[t]
	\includegraphics[width=.65\linewidth]{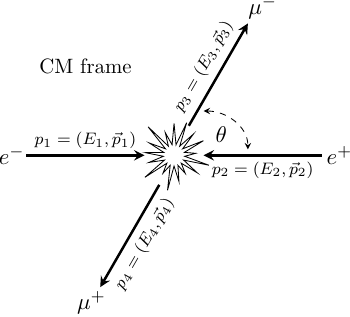} 
	\caption[Scattering in the centre-of-mass frame]{Kinematics of $e^- e^+ \rightarrow \mu^- \mu^+$ in the centre-of-mass frame. Since the particles are all on-shell, $p_i = \sqrt{E^2 - m_e^2}$ and $p_f = \sqrt{E^2 - m_\mu^2}$.}
	\label{fig:CMscattering}
\end{figure}

In this way, we can impose on-shell conditions through the use of these variables. In our specific case, they read
\begin{align}
	\label{eq:OnShell1}
	s + t + u = 2 m_e^2 + 2 m_\mu^2  ,
\end{align}
which always holds for asymptotic states and 
\begin{align}
	\label{eq:OnShell2}
	\Lambda = 4 \bar R \, ,
\end{align}
where $\bar R$ is the curvature of the background metric. In our computation, we have used the flat Minkowski metric and therefore $\Lambda = 0$. This last condition was dictated by the fact that all computations have been performed in an expansion around a flat Minkowski metric. The same approximation for the on-shell conditions has been used in \cite{Knorr:2022lzn} as well. Once gauge independence has been verified, we can fix the kinematics by neglecting the fermion masses given that $\sqrt{s} \gg m_e, m_\mu$. At this stage, we are left with the following matrix element squared in terms of the Mandelstam variables
\begin{align}
	\langle \vert \mathcal{M}_{fi} \vert^2 \rangle =&\\[1ex]
	&\hspace{-1.5cm}\frac{\pi^2 G_{\text{N}}^2}{s^2} \left[s^4-4 s^2 (t-u)^2+(t-u)^2 \left(5 t^2-6 t u+5 u^2\right)\right].\notag
\end{align}
In the relativistic case, $t$ and $u$ can be expressed in terms of the centre-of-mass energy squared $s$ and the scattering angle $\theta$ as shown in \cref{fig:CMscattering}. This allows us to write the scattering amplitude according to
\begin{align}
	\label{eq:SAurl}
	\langle \vert \mathcal{M}_{fi} \vert^2 \rangle = 
	\pi^2 s^2 G_{\text{N}}^2 \left( 1 - 3 \cos^2 \theta + 4 \cos^4 \theta \right) \, .
\end{align}
By using the well-known formula for the differential cross section in the centre-of-mass frame
\begin{align}
	\label{eq:dsCM2}
	\left. \frac{\mathrm d\sigma}{\mathrm d\Omega} \right|_{\rm CM} = \frac{1}{64 \pi^2 s} \frac{p_f}{p_i} \langle \vert \mathcal{M}_{fi} \vert^2 \rangle \,  \theta \left(\sqrt{s} - m_3 - m_4\right)  ,
\end{align}
and integrating over the solid angle $\Omega$, we obtain the following formula for the leading-order graviton-mediated cross-section
\begin{align}
	\sigma_h = \frac{\pi}{20}  s \, G_{\text{N}}^2\,.
\end{align}

\section{Reconstruction of the graviton spectral function}
\label{app:reconstruction}

Here, we only focus on the spectral function of the traceless-transverse part $\mathcal{G}_{hh}$ of the graviton propagator in a flat background. The leading asymptotics of $\mathcal{G}_{hh}(p)$ are proportional to $1/p^2$ in the infrared and $p^{\eta_h - 2}$ in the ultraviolet, with $\eta_h \approx 1.03$~\cite{Bonanno:2021squ}. The asymptotic of $1/p^2$ captures the classical IR regime, namely we obtain the classical gravity described by the Einstein-Hilbert action. This term contributes a Dirac delta for vanishing frequencies in the spectral density. Since we know the dominant contribution in the infrared analytically, we exclude it from the reconstruction. In this regard, we focus on reconstructing the difference propagator $\Delta \mathcal{G}_{hh}^{(1)}$ defined by
\begin{align}
	\label{eq:DeltaG1}
	\Delta \mathcal{G}_{hh}^{(1)} (p) = \mathcal{G}_{hh} (p) - \frac{1}{p^2} \, .
\end{align}
The latter quantity presents, like the $1/p^2$ pole, a divergence in the infrared but a subleading log-like one. The method we are going to use, also known as the Schlessinger Point Method (SPM)~\cite{PhysRev.167.1411}, fails in reproducing logarithmic divergences just as it fails in reproducing divergences of the $1/p^2$ type. 
An ideal reconstruction is based on the use of analytic fits. These in the IR (UV) must not interfere with the UV (IR) behaviour and must not introduce further structures. For example, in the case of~\cref{eq:DeltaG1}, the subtraction with $1/p^2$ satisfies this property. This structure is dominant in the IR but is suppressed for high values of the momentum due to $p^{\eta_h -2}$. 
Following the same \textit{modus operandi}, we should use an analytic function that behaves like a logarithm for small momenta to reproduce the log-like divergence. At the same time, it must decrease more rapidly in the UV than a usual logarithm in order not to affect the asymptotic behaviour. 
After several attempts with various analytical structures, we have concluded that the best one for this purpose is the same used in~\cite{Bonanno:2021squ}, i.e., the confluent hypergeometric function $\mathcal{U}_{a,b} (p^2)$, whose leading large-momentum asymptotic is $1/p^{2a}$. For $b =1$ and for small momenta, it reads
\begin{align}
	\lim_{p \rightarrow 0} \mathcal{U}_{a,1}(p^2) = - \frac{1}{\Gamma(a)} \left( 2 \gamma_{\rm E} + \frac{\Gamma^{\prime}(a)}{\Gamma(a)} + \text{ln}(p^2) \right) \, ,
\end{align}
where $\gamma_{\rm E}$ is the Euler-Mascheroni constant and $\Gamma (z)$ is the gamma function. Note that we are implementing dimensionless momenta $p^2 \rightarrow p^2 / M_{\text{Pl}}^2$ and dimensionless propagator and spectral function
\begin{align}
	\mathcal{G}_{hh} &\rightarrow M_{\text{Pl}}^2 \, \mathcal{G}_{hh} ,
	&
	\rho_h &\rightarrow M_{\text{Pl}}^2 \, \rho_h \, .
\end{align}
The confluent hypergeometric function does not introduce any poles in the positive real half-plane and it is UV subleading for $a > 1 - \eta_h /2 \approx 0.49$. These features make it the perfect candidate for the reconstruction of 
\begin{align}
	\label{eq:DeltaG2}
	\Delta \mathcal{G}_{hh}^{(2)} (p) = \Delta \mathcal{G}_{hh}^{(1)} (p) - A_h \, \mathcal{U}_{1,1}(p^2) \, ,
\end{align}
where 
\begin{align}
	\mathcal{U}_{1,1} (p^2) = e^{p^2} \Gamma(0, p^2) \, ,
\end{align}
with the upper incomplete gamma function 
\begin{align}
	\Gamma (a, z) = \int_{z}^{\infty} \mathrm dt \, t^{a-1} \, e ^{-t} \, .
\end{align}
In conclusion,~\cref{eq:DeltaG1,eq:DeltaG2} subtractions in the infrared leave us with a constant contribution that remains for small momenta
\begin{align}
	\lim_{p \rightarrow 0} \Delta \mathcal{G}_{hh}^{(2)}(p) \approx 0.29 \, .
\end{align}
Now there are no more divergences for small values of the momenta. 
By applying the SPM, we are able to replicate the asymptotic behaviour in the ultraviolet and also the convergent behaviour of $\Delta \mathcal{G}_{hh}^{(2)}$ in the infrared. The algorithm for the SPM fit is given in~\cref{app:SPM}. By adding the confluent hypergeometric function $\mathcal{U}_{1,1} (p^2)$  to the present SPM fit, we can reproduce $\Delta \mathcal{G}_{hh}^{(1)}$. What remains to be done is only the determination of $A_h$, see~\cref{eq:DeltaG2}. In this work, we have obtained $A_h \approx 0.11$, which agrees with the numerical value found in~\cite{Bonanno:2021squ}. 

Thus, in order to obtain the whole fluctuation propagator $\mathcal{G}_{hh}$, it would be sufficient to add $1/p^2$ as well. However, we aim to reconstruct the continuous part of the spectral function. Therefore, we can ignore this contribution which we know would provide a Dirac delta with its centre at zero frequency. To simplify things, we can parameterise the entire spectral function as follows
\begin{align}
	\rho_h (\omega) = \frac{\delta (\omega)}{\omega} + \rho_{h}^{\text{cont}} (\omega) \, ,
\end{align}
where the Dirac delta comes from the $1/p^2$ in the Euclidean propagator and $\rho_{h}^{\text{cont}}$ stems from $\Delta \mathcal{G}_{hh}^{(1)}$. Therefore, the Euclidean propagator reads in the following way
\begin{align}
	\mathcal{G}_{E} (p) = \frac{1}{p^2} + A_h \mathcal{U}_{1,1} (p^2) + \Delta \mathcal{G}_{hh}^{(2), \text{SPM}} (p) \, .
\end{align}
We can perform a Wick rotation of the last two quantities in the right-hand side of the previous equation and then by taking their imaginary part, see~\cref{eq:ImGE}, we obtain the continuous part of the spectral density $\rho_{h}^{\text{cont}}$ displayed in the right-top corner of~\cref{fig:spectral-function} on the left. 

\begin{figure}[t]
	\includegraphics[width= \columnwidth]{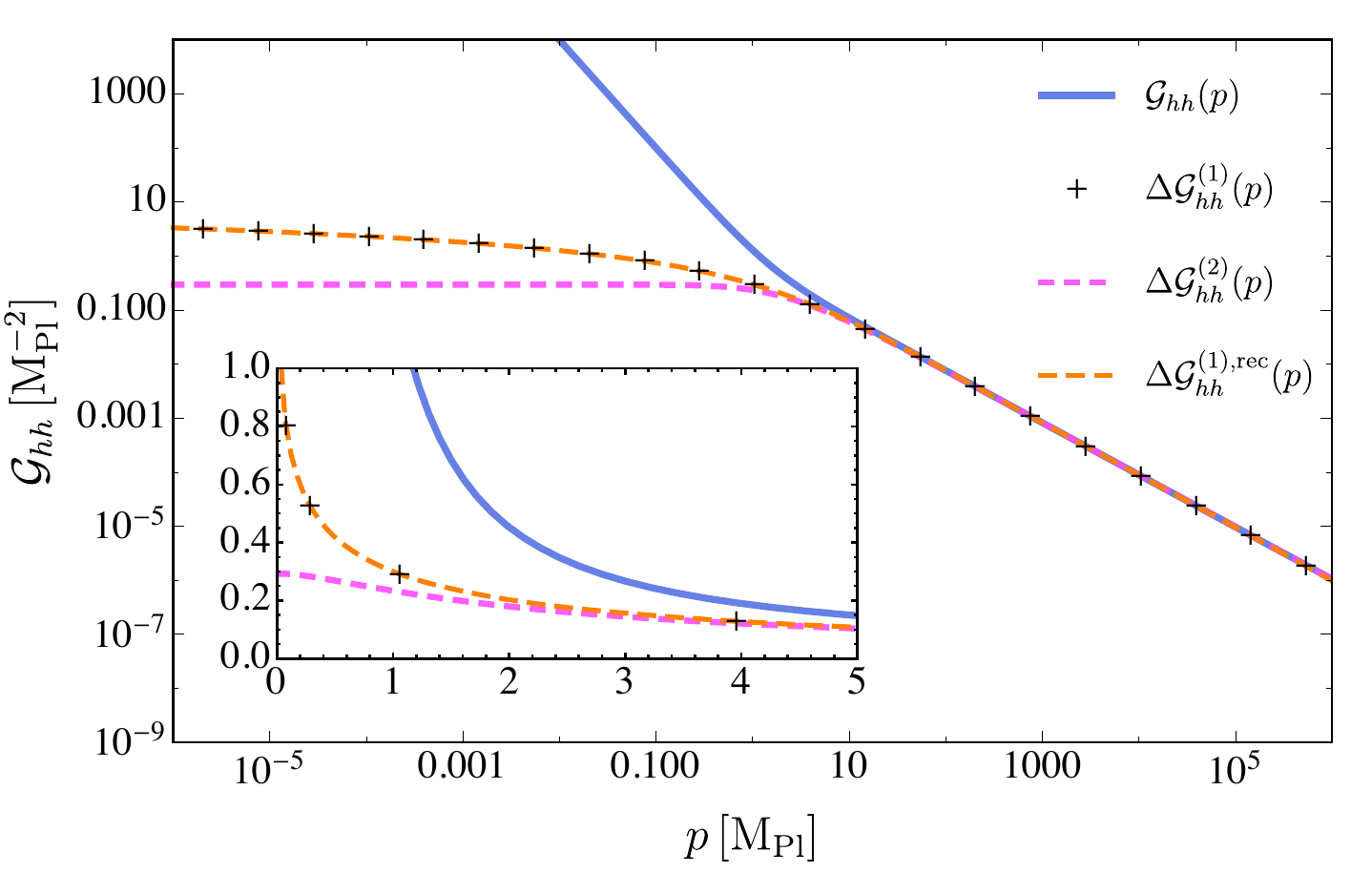} 
	\caption{Comparison of $\Delta \mathcal{G}_{hh}^{(1), \rm rec}$ (dashed orange line) reconstructed in this work with that in~\cite{Bonanno:2021squ}, $\Delta \mathcal{G}_{hh}^{(1)}$ (black plus markers). The difference propagator $\Delta \mathcal{G}_{hh}^{(1), \rm rec}$ has been computed by plugging $\rho_h^{\text{cont}}$ into the spectral representation \cref{eq:SpecIntEucl}.}
	\label{fig:SPMep}	
\end{figure}

In order to compare the spectral function reconstructed in this work -- $\rho_{h}^{\text{cont}}$ -- with the first one in~\cite{Bonanno:2021squ}, denoted by $\rho_{h}^{\text{rec}}$, we checked the IR asymptotics and the percentage error.
Here, we obtain for $\rho_{h}^{\text{cont}}$ the same asymptotic value for zero frequency that is shown in~\cite{Bonanno:2021squ} 
\begin{align}
	\lim_{\omega \rightarrow 0} \rho_{h}^{\text{cont}} (\omega) \approx 0.7 \, .
\end{align}
The percentage error has been estimated according to
\begin{align}
	E_{\%, i} = 100 \cdot \left\vert \frac{\rho_{h}^{\text{cont}}(\omega_i) -\rho_{h}^{\text{rec}}(\omega_i)}{\rho_{h}^{\text{cont}}(\omega_i)} \right\vert \, ,
\end{align}
where the index $i$ runs over the sampled data points considered for the fit. To verify the consistency of such a reconstruction, we must expect to obtain again the difference propagator $\Delta \mathcal{G}_{hh}^{(1)}$ from the spectral integral~\cref{eq:SpecIntEucl} with $\vert \vec{p} \, \vert = 0$. This last check is shown in~\Cref{fig:SPMep}. We note a good agreement with the original Euclidean data, supported by a hasty error analysis. For each sampled point, the estimated percentage error 
\begin{align}
	E_{\%, i} = 100 \cdot \left\vert \frac{\Delta \mathcal{G}_{hh}^{(1)} (p_i) -\Delta \mathcal{G}_{hh}^{(1), \text{rec}} (p_i)}{\Delta \mathcal{G}_{hh}^{(1)} (p_i)} \right\vert \, ,
\end{align}
is always below $10 \%$, which supports that the reconstruction occurred efficiently.

\subsection{Schlessinger point method}
\label{app:SPM}
The Schlessinger Point Method, also known as Resonances Via Pad{\'e} (RVP) method, is based on a rational-fraction representation similar to Pad{\'e} approximation methods. The rational-fraction construction interpolates a set of N points $(x_i, y_i)$ such that
\begin{align}
	C_{\text{N}} (x) = \dfrac{y_1}{1+\dfrac{a_1 (x - x_1)}{1+ \dfrac{a_2 (x - x_2)}{\vdots \dots a_{\text{N}-1} (x - x_{\text{N}-1})}}}  \, .
\end{align}
It is immediately apparent that $C_{\text{N}} (x_1) = y_1$ and that the coefficients $a_{1}, a_{2}, \ldots, a_{\text{N}-1}$ are chosen in order to get $C_{\text{N}} (x_i) = y_i \, \, \forall \, i$. They are determined by using a recursive formula that applies for every $a_i$ but $a_1$. For the latter we have
\begin{align}
	a_1 = \frac{\left( y_1 / y_2 \right) - 1}{x_2 - x_1} \, ,
\end{align}
and in general
\begin{align}
	a_l = \frac{1}{x_l - x_{l+1}} \left\{ 1 + \dfrac{a_{l-1}(x_{l+1} - x_{l-1})}{1 + \dfrac{a_{l-2}(x_{l+1} - x_{l-2})}{\vdots \dots \dfrac{a_1 (x_{l+1} - x_1)}{1 - (y_1 / y_{l+1})}}} \right\} \, .
\end{align}

\bibliography{bibliography.bib}

\end{document}